\documentclass[3p,times]{elsarticle}
\DeclareGraphicsExtensions{.pdf,.gif,.jpg,.eps,.png} 

\usepackage{amssymb,amsmath} 
\usepackage{graphicx}  
\usepackage{natbib}   
\usepackage[colorlinks=true]{hyper ref}  
\usepackage{xcolor}
\usepackage{stackengine}
\usepackage{color,soul}
\usepackage{booktabs}
\soulregister\cite7
\soulregister\ref7
\soulregister\pageref7

\usepackage[title]{appendix}

\def\Put(#1,#2)#3{\leavevmode\makebox(0,0){\put(#1,#2){#3}}}

\usepackage{mathabx}
\definecolor{Mblue}{rgb}{0,    0.4470,    0.7410}
\definecolor{Mred}{rgb}{0.8500,    0.3250,    0.0980}
\definecolor{Myellow}{rgb}{0.9290,    0.6940,    0.1250}
\definecolor{Mpurple}{rgb}{0.4940,    0.1840,    0.5560}
\definecolor{Mgreen}{rgb}{0.4660,    0.6740,    0.1880}
\definecolor{MColor6}{rgb}{0.3010,    0.7450,    0.9330}
\definecolor{MColor7}{rgb}{0.6350,    0.0780,    0.1840}
\definecolor{Nyellow}{rgb}{0.9290,    0.6940,    0.0250}
\definecolor{Nyellow}{rgb}{1,    0.80,    0.00}
\usepackage{booktabs}

\sethlcolor{white}

\makeatletter
\def\ps@pprintTitle{%
   \let\@oddhead\@empty
   \let\@evenhead\@empty
   \let\@oddfoot\@empty
   \let\@evenfoot\@oddfoot
}
\makeatother

\def\p{\partial} 
\def\ub{\mathbf{u}}
\newcommand{\bV}{{\bf V}}
\newcommand{\bU}{{\bf u}}

\newcommand{\divergence}{{\rm div}}
\newcommand{\odivergence}{{\rm odiv}}
\newcommand{\grad}{{\rm grad}}

\def\ub{\mathbf{u}}

\def\xb{\mathbf{x}}

\newcommand*{\nolink}[1]{%
  {\protect\NoHyper#1\protect\endNoHyper}%
}

\newcommand{\addtxt}[1] {\hl{#1}}
\usepackage[capitalise,nameinlink]{cleveref}

\begin{document}
\begin{frontmatter}
	\title{Numerical simulation of compressible flows\\ using two-phase observable Euler equations}
	\author[UF]{Bahman Aboulhasanzadeh \corref{cor1}}
	\cortext[cor1]{Corresponding author. \textit{Email address:} baboulha@alumni.nd.edu}
	\author[UF]{Kamran Mohseni \corref{cor2}}
	\cortext[cor2]{Corresponding author. \textit{Email address:} mohseni@ufl.edu. \textit{Tel.:} +1 352 273 1834}

	\address[UF]{Department of Mechanical \& Aerospace Engineering, University of Florida, Gainesville, FL 32611, USA}
	\date{}

	\begin{abstract}
		Most fluid flow problems that are vital in engineering applications involve at least one of the following
		features: turbulence, shocks, and/or material interfaces. While seemingly different phenomena, these flows all
		share continuous generation of high wavenumber modes, which we term the `$k_\infty$ irregularity.' In this work,
		an inviscid regularization technique called `observable regularization' is proposed for the simulation of two-phase compressible flows. The proposed approach regularizes the equations at the level of the partial
		differential equation and as a result, any numerical method can be used to solve the system of equations. The
		regularization is accomplished by introducing an `observability limit' that represents the length scale below
		which one cannot properly model or continue to resolve flow structures. \hl{An observable volume fraction equation is derived for capturing the material interface, which satisfies the pressure equilibrium at the interface.} The efficacy of the observable
		regularization method is demonstrated using several test cases, including a one-dimensional material interface
		tracking, one-dimensional shock-tube and shock-bubble problems, and two-dimensional simulations of a shock
		interacting with a cylindrical bubble. The results show favorable agreement, both qualitatively and
		quantitatively, with available exact solutions or numerical and experimental data from the literature. 
		The computational saving by using the current method is
		estimated to be about one order of magnitude in two-dimensional computations and significantly higher in three-dimensional computations. Lastly, the effect of the observability limit and best practices to choose its value
		are discussed.
	\end{abstract}


\end{frontmatter}
\section{Introduction}

Material interfaces and/or shockwaves are the source of complex flow behavior in many fluid engineering
problems. Scientific understanding of these problems is the base for technological advances needed in engineering and
daily life. For example, shock-induced collapse is a damaging phenomena leading to propeller erosion in naval
engineering \cite{KnappRT:70a}; similar mechanisms may lead to tissue damage and internal bleeding in shockwave
lithotripsy \cite{SturtevantB:01a,SapozhnikovO:02a,JohnsenE:07a}, a non-invasive medical treatment of kidney
stones. Understanding such phenomena can help us to develop designs that specifically avoid cavitation or control it to
our advantage \cite{BrennenC:11a}. In the past couple of decades, with the improvement of computational power,
numerical study of such problems has become mainstream. While solving the conservation equations, discontinuity in flow
variables creates challenges that need specific attention.

In the past three decades, significant effort has been given to numerical treatment of material
interfaces and shocks. Many approaches are developed for tracking/capturing interfaces. In Front Tracking
\citep{ChernIL:86a}, a separate unstructured grid is used to follow the location of the interface, and although it has
many advantages when it comes to control over the topology change and sharpness of the interface, it has more complexity
compared to other methods and also is challenging for parallel processing. Eulerian interface tracking approaches have
lower complexities in maintaining and parallel processing of the interface. However, they still need reconstruction of
the interface, which adds to computational cost of method. Examples of such methods are ghost-fluid method (GFM) and
volume-of-fluid methods. In GFM, a level set function is introduced with a zero value representing the interface. The
equation for each fluid is solved on each side of the interface with the help of ``ghost-fluid'' points that are
reconstructed across the interface at each time step \cite{Osher:99a,XuS:97a}. Volume of fluid methods keep the
interface sharp by reconstructing the interface from the advected volume fraction \cite{ShyueKM:06a}. Front capturing
methods are the simplest type for multiphase flow. These methods solve an additional equation for evolving the interface
without the need for interface reconstruction. The pressure non-equilibrium effect at the interface was initially a
downside of these methods as a result of numerical diffusion. Karni \citep{KarniS:94a} first addressed this issue
by solving the governing equations in non-conservative form and solving an advection equation for the mass fraction
of one of the phases. Many other methods were later developed to address this issue and each solves an additional
advection equation for either a material property \cite{AbgrallR:96a,ShyueKM:98a}, mass fraction \citep{KarniS:96a}, or
volume fraction \citep{KokhS:02a}. In addition to development of methods for tracking/capturing the interface,
several Riemann solver type approaches are developed for physical approximation of fluxes at discontinuities,
specifically for problems including shock waves; Roe solver \citep{RoePL:81a}, HLL \citep{HartenA:83b}, HLLC
\citep{ToroE:94a}, and flux vector splitting \citep{StegerJL:81a} are a small sample of such methods. All the efforts
in solving fluid equations are based on the \textit{assumption} that any set of conservation equations
(e.g. Euler) are valid when solved with finite resolution. This assumption holds in regions of smooth variations
where the flow variables and their derivatives can be represented using the limited resolution with minimal
error. However, this may not be the case in regions with sharp changes.

When there are sharp variations in a field quantity, one needs to have a large number of wave modes in the
Fourier space to correctly represent the data, and as the thickness of the jump decreases the required number of modes
approaches infinity. \hl{Considering that discontinuous solutions are part of admissible solutions of Euler equations; this infinite resolution requirement is inherent in the Euler equations before any discretization scheme imposed. Consequently, it should not be a surprise that naive discretization of Euler equations could lead to numerical instability} in simulation of shocks, interfaces, and even turbulence. In order to avoid this conflict, \citet{Mohseni:09w, Mohseni:10w} developed the concept of {\it observable divergence} in which the finite resolution
limit (observability limit) is introduced into the derivation of conservation equations; for a derivation of the
observable divergence theorem see \cite{Mohseni:10w}. \citet{ZhangT:14a} demonstrated that observable divergence can be used to efficiently regularize the relativistic Burgers equation. They showed that the regularized equation preserves the shock speed and conservation properties of the original equation. \hl{It is believed that observable Euler equations do not allow discontinuous solutions; as a result of this regularity, these equations are expected to have better numerical behavior while handling shocks and interfaces.} Using observable divergence for conservation laws requires a
reevaluation of the governing equations for the existing problems in order to systematically resolve the challenges in
the simulations of fluid flows, including but not limited to multiphase flows, shocks, and turbulence.

In the observable approach the regularization is done at the level of the partial differential equation. The aim of
this paper is to numerically demonstrate that the governing equations derived using this concept regularize shocks and interfaces
with no need for special numerical \hl{interface} treatment. To this end, we use pseudo-spectral method which has zero numerical
viscosity. \hl{Note that the selection of pseudo-spectral method is just to demonstrate that even for a numerical scheme with no numerical dissipation the observable equations are regularized. One can easily use any other numerical scheme. We have also used finite difference and high order Pad\'e schemes in the past.} The regularizing capability of the observable set of equations is demonstrated with some examples. We first
develop the observable volume fraction equation for a two-phase flow problem. Then, present the numerical solution of 1D and 2D
shock-interface interaction problems using the observable set of equations and compare them to available exact
solutions or experimental/numerical results from literature. 
The effect of observability limit on the results is investigated at the end and a guideline for selecting
the observability limit is presented.

\section{Motivation and Observability Concept}
\label{section:observable-concept}
In physics, a \emph{field} is a physical quantity that can be assigned a value at any and each mathematical point in space and time; in other words a field is assumed to be a continuum. 
In this section, a critical review of some of the basic concepts and axiomatic assumptions in classical field theory is presented. We further show the intimate connection between these assumptions and many of the most well known, intractable challenges in classical physics such as turbulence and high dimensionality problems in shocks and multiphase flows. The main difficulties in these problems stem from the requirement of resolving fields to infinitely-small scales while in reality no experiments or simulations are capable of such resolution. A resulting consequence is what we refer to as $k_\infty$ (wavenumber infinity) irregularity; see the discussion below. Our proposed solution to these issues is to introduce an \emph{observable} field quantity. Equipped with this new observability concept, we have systematically revisited some of the generally accepted concepts from the calculus and show that one can derive governing observable dynamical equations which are a more realistic representation of the observable physical world. To explain our ideas, in the following we consider three seemingly different phenomenon from fluid mechanics; namely shocks, turbulence, and sharp interfaces in two-phase flows. We will argue that these different challenges are, in fact, manifestations of the infinite observability assumption in the definition of a field quantity. We will then offer a remedy.

\begin{figure}[htbp]
\vspace{-0mm}
\begin{center}
  \includegraphics[width=.65\linewidth]{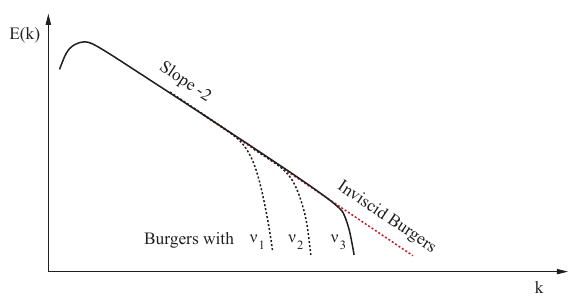}
\caption{ The $k_\infty$ irregularity in a field quantity. The schematics show the energy spectrum in a field quantity at long time. } \label{fig:HighWavenumber}\vspace{-8mm}
\end{center}
\end{figure}
\medskip
\noindent \emph{$k_\infty$ Irregularity. } 
To introduce the problem of assuming infinite observability, consider the simplest and standard example of shock formation in the inviscid Burgers' equation, $u_t+uu_x = 0$. It is well known that a smooth initial condition for the Burgers' equation can generate a finite-time discontinuity.
Formation of a discontinuity is characterized by a rather fast generation of increasingly higher wavemodes as the solution profile steepens toward a shock; see Figure \ref{fig:HighWavenumber}. For the one-dimensional Burgers equation the cascade of energy to smaller scales has a slope of $-2$.
The conventional wisdom to remedy nonphysical discontinuities has been the addition of a viscous term, e.g. in the form of a Laplacian of the velocity field, in order to control and reduce the creation of smaller scales. The Laplacian term changes the nature of the partial differential equation (PDE) from a hyperbolic PDE to a regularized parabolic one. Note that the term responsible for this high-wavenumber generation is the $\ub \cdot \nabla \ub$ term which is the difference between an Eulerian and Lagrangian observer; we have termed this phenomenon the \emph{high wavenumber} ($k_\infty$) \emph{irregularity}. 

The $k_\infty$ irregularity is a common feature of almost all challenging problems in fluid dynamics including turbulence, shocks and sharp interface two-phase problems where one requires comparing observations between Eulerian and Lagrangian observers ($\ub \cdot \nabla \ub$ term). The symptoms of the problem in each case (shocks, turbulence, two-phase flows) are identical; namely a never ending tail of the Fourier spectrum with the signal in the physical domain showing formation of a discontinuity or excessive filamentation. Again, the \emph{ad hoc} remedy in each case has been the introduction of a parabolic term in the form of a Laplacian or its higher order versions. To this end, \emph{the $k_\infty$ irregularity is responsible for excessive computational cost of high Reynolds number flows, high Mach number shocks, and high density ratio two-phase flows. Interestingly, this $k_\infty$ irregularity has not been recognized among these seemingly different problems and as a result different remedies were considered in different disciplines.}

Now, a fundamental question is: at which step in deriving the governing PDEs in classical field theory were nonphysical solutions allowed to creep in as a part of the domain of admissible solutions to a physical problem? We have recently argued that the answer to this question is in the \emph{infinite observability assumption} that is automatically accepted when a physical quantity is represented as a field. In other words, the values of the quantity at any two points can be distinguished from each other no matter how small the distance between the two are! Notice that the wavenumber $k \rightarrow \infty$ is the same as the length scale $L\rightarrow 0$. \emph{Basically, by merely defining a field quantity we have axiomatically accepted this hidden assumption}. As a result, we have made the assumption that any length scale, no matter how small, is observable by a field quantity. It is clear then that this is not a physically correct assumption. No field quantity can be experimentally or numerically observed to an infinitely-small length scale. This physical limit in our observation is termed the \emph{observability scale}, $\alpha$. Anything below this scale is virtually unmeasurable or unobservable from a physical stand point. As a result, \emph{what we often observe is an averaged quantity rather than the mathematical limit where the observation volume approaches zero. This has significant ramifications when one calculates the flux of a quantity at the face of a volume surrounding a point of interest}. This limit is \emph{not} reflected in the classical field theory equations and we believe that it is the source of many sustained theoretical and numerical difficulties in solving some of the challenging problems in classical field theory, including fluid mechanics.

Upon asking this question a few years ago, we were led to ask a series of others: What would happen if we start with a field quantity with \hl{a finite} observability limit? Do we obtain the same governing dynamical equations? Do we still obtain the same integral equations in calculus (e.g., divergence and Stokes' theorems) if the derivations are performed with an observable field quantity and not, as is currently done, with an infinity observable quantity? The following provides answers to some of these questions; see \cite{Mohseni:09w, Mohseni:10w} for details.

\noindent{\it Observable Field Quantity}. An observability limit on a field quantity, $f$, could be imposed by a convolution kernel, $g$, with a length scale $\alpha$ to obtain 
\begin{equation}
	\overline{f} = g \ast f
	\label{eqn:observable_quantity}
\end{equation} 
where $\ast$ is the convolution operator. Here, $f$ is the infinitely observable quantity as defined in all classical field theory and $\overline{f}$ is the observable field quantity. The following theorem addresses how the classical divergence theorem is altered for observable field quantities:

\medskip
\noindent \emph{Observable Divergence Theorem} \cite{Mohseni:09w,Mohseni:10w}. Let $\Omega$ be a region with surface boundary $S$ oriented outward. Let $\mathbf{F}(\mathbf{x}) = f \mathbf{V}(\mathbf{x}) =  (u(\mathbf{x}), v(\mathbf{x}), w(\mathbf{x}) ) f$ be a vector function that is continuous and has continuous first partial derivatives in a domain containing $\Omega$. Then\vspace{-2mm}
\begin{equation}
  \iiint\limits_\Omega \odivergence\, \mathbf{F} =  \iiint\limits_\Omega \left( \overline{f}\, \divergence\mathbf{V} + \overline{\mathbf{V}} \cdot \grad f \right) \; dV 
  = \iint\limits_S \mathbf{F \cdot n} \; dS
  \label{Eqn:observable_divergence}
\end{equation}
where the operator $\bar{(\cdot)}$ is defined by $\bar{(\cdot)} = g \ast (\cdot)$, $\ast$ is the convolution operator, $g$ is the kernel of the convolution with an observability length scale $\alpha$. $g^\alpha$ could be any kernel with properties stated in \Cref{FilterProperties}.

\begin{table}[htbp]
\caption{\label{FilterProperties} Requirements for the averaging kernel.}
\begin{center}
\begin{tabular}{@{}lc@{}}
\toprule
Properties & Mathematical Expression\\
\midrule
Normalized & $\int g$=1 \\
Nonnegative & $g(\xb) >0,\, \forall \xb $   \\
Decreasing & $|\xb_1|\ge |\xb_2| \Rightarrow g(\xb_1) \le g(\xb_2)$  \\
Symmetric  & $|\xb_1|= |\xb_2| \Rightarrow g(\xb_1) = g(\xb_2)$    \\
\bottomrule
\end{tabular}
\end{center}
\end{table}

\noindent \hl{See \cite{Mohseni:09w,Mohseni:10w} for a proof.}
	
\hl{Observable method was systematically evaluated numerically and analytically for multidimensional Burgers equations} \cite{Mohseni:09h,Holm:03a}. \citet{Mohseni:09h} proved the existence and uniqueness of the
solution to observable Burgers' equation and its convergence to the weak solution of Burgers' equation as $\alpha \to 0$ for a subset of kernels with
the form $\hat g(k) = 1/\left( {1 + \sum\nolimits_{j = 1}^N {{C_j}{k^{2j}}} } \right)$; a hat represents the Fourier
transform of a quantity and the $C_j$ are the kernel constants. Furthermore, the observable Burgers' equation is a
special case of a more general equation, which is shown to be Hamiltonian by 
\citet{Holm:03a}. 

Equipped with the observable divergence theorem one can systematically derive the observable equations for any conservation laws in classical field theory. In the following section, the conservation
equations for a compressible flow are presented first and then an observable volume fraction equation is derived for capturing the material interface
and defining material properties inside each phase.

\section{Governing Equations for Two-phase Flow}
\noindent {\it Infinitely observable Euler equations}. Conservation of mass, momentum, and energy for
the mixture fluid properties are inferred in different ways
\citep{KokhS:02a,MurroneA:05a} from Baer-Nunziato model \citep{BaerM:86a} or Drew derivation of two phase
conservation equations \cite{DrewDA:71a,DrewD:83a}:
\begin{align}
	 & \frac{{\partial \rho }}{{\partial t}} + \,\nabla  \cdot \rho \bold{u}\, = 0, \label{eqn:EulerMass}                                \\
	 & \frac{{\partial \rho \bold{u}}}{{\partial t}} + \,\nabla  \cdot \rho \bold{u}\bold{u}\, + \nabla p = 0, \label{eqn:EulerMomentum} \\
	 & \frac{{\partial \rho E}}{{\partial t}} + \,\nabla  \cdot \left( {\rho E\bold{u} + p\bold{u}} \right) = \,0,
	\label{eqn:EulerEnergy}
\end{align}
where $\rho$, $\bold{u}$, $p$, and $E$ are density, velocity vector, pressure and specific total energy of fluid
mixture, respectively.  In the derivation of \cref{eqn:EulerMass,eqn:EulerMomentum,eqn:EulerEnergy}, an instantaneous
mechanical relaxation between phases is assumed, $u=u_1=u_2$ and $p=p_1=p_2$ where subscript 1 and 2 refer to phase 1
and 2. The mixture quantities are defined using the following relations:
\begin{equation}
	\rho  = {\rho _1}{z _1} + {\rho _2}{z _2},\quad\quad\quad\quad\quad \rho E = {\rho _1}{E_1}{z _1} + {\rho _2}{E_2}{z _2},
\end{equation}
here, $z_1$ and $z_2$ are the volume fractions of phase 1 and 2 and $z_1 + z_2 = 1$. In the rest of this text $z$ with
no subscript refers to $z_1$.

\noindent {\it Observable Euler equations}. \citet{Mohseni:09w,Mohseni:10w} derived the observable mass, momentum, and energy equations. Here, we repeat the derivation for the sake of completeness.  We develop the differential equations that must be satisfied by a fluid with an observability limit of $\alpha$. The equations are expected to resolve all relevant dynamical quantities up to this {\it observable} scale $\alpha$. As mentioned before, this scale is dictated by the resolution of a particular numerical simulation or the length scale resolution of a particular experimental equipment. For scales below $\alpha$ we assume that the medium still acts as a continuum and one can take the mathematical limit of $\Delta V \rightarrow 0$, where $\Delta V$ is the control volume of interest. However, our {\it observable} scale remains at $\alpha$ even as this mathematical limit is taken to zero.  

Before the differential form of the  observable conservation laws for mass, momentum, and energy of a continuum are presented, we will formulate the general form of the observable differential conservation law for a quantity per unit volume, $f$. We assume $f$ satisfies a conservation law of the form\vspace{-2mm}
\begin{equation}
 \textrm{Rate of generation of $f$} = Q\vspace{-2mm}
\end{equation}
where $Q$ is the sum of all sources of $f$ inside the volume. For instance $Q$ is the total external force exerted on a control volume if one considers the conservation of momentum inside the volume. The rate of generation of $f$ is basically the outflow of $f$ minus the inflow at the boundaries of the volume plus the storage inside the volume. The rate of outflow minus the inflow at the boundaries can be conveniently calculated by the operator $\odivergence \,f \bV$ where $\bV$ is the continuum velocity, see \Cref{eqn:observable_quantity}. Therefore, the observable conservation of $f$ reduces to the following differential equation\vspace{-2mm}
\begin{equation}
  \dfrac{\p f}{\p t} + \overline{f}\ \divergence \bV + \overline{\bV} \cdot \grad f = Q.\vspace{-2mm}
  \label{eq:Of1}
\end{equation}

\noindent Now, the observable compressible Euler equations can be easily derived by using the conservation of mass ($\rho$), momentum ($\rho \mathbf{u}$), and energy ($\rho E = \rho(e + \frac{1}{2} u^2)$) to obtain
\begin{subequations}\label{eq:OEuler1}
\begin{align}
  &\dfrac{\p \rho}{\p t} + \overline{\rho}\ \nabla \cdot \bU + \overline{\bU} \cdot \nabla \rho = 0.
  \label{eq:OEulerMass1} \\
  &\dfrac{\p \rho \bU}{\p t} + \overline{\rho\bU} \ \nabla \cdot \bU + \overline \bU \cdot \nabla (\rho \bU) = -\nabla p.
  \label{eq:OEulerMom1}\\
  &\dfrac{\p \rho E}{\p t} + \overline{ \rho E} \ \nabla \cdot \bU + \overline{\bU} \cdot \nabla (\rho E)  =  
   - \left( \overline{p}\ \nabla \cdot \bU + \overline{\bU} \cdot \nabla p \right) + \rho \bU \cdot \mathbf{\Sigma} + S,
  \label{eq:OEulerEn1}
\end{align}
\end{subequations}
where $\mathbf{\Sigma}$ is the sum of all external forces acting on the control volume or its surface and $S$ is the sum of all energy sources inside the control volume. Note that in the limit of $\alpha$ approaching zero, that is when the observable scale approaches a mathematical zero, one recovers the classical Euler equations. For the problems presented in this work $\mathbf{\Sigma}$ and $S$ is assumed to be zero.

In order to close this system of equations, an equation of state (EOS) needs to be used. Here, we use the ideal gas
equation of state which is used by many authors for the gas-gas problems \cite{KarniS:96a,AbgrallR:96a}:
\begin{equation}
	\rho E - \frac{1}{2}\rho {\bold{u}^2} = \Gamma p,\quad\quad\quad\quad\quad \Gamma  = \frac{1}{{\gamma  - 1}},
\end{equation}
where $\gamma$ is the ratio of specific heat of material at constant pressure to its specific heat at constant
volume. The mixture $\Gamma$ is defined using:
\begin{equation}
	\Gamma  = \frac{1}{{\gamma  - 1}} = \frac{{{z _1}}}{{{\gamma _1} - 1}} + \frac{{{z _2}}}{{{\gamma _2} - 1}}.
	\label{eqn:mixturematerialprop}
\end{equation}
\subsection{Observable Interface Capturing}
\hl{To apply the observable} \Cref{eq:OEulerMass1,eq:OEulerMom1,eq:OEulerEn1} \hl{to two-phase  flows, we need to also obtain an equation for capturing the interface, which is needed to define material properties in each phase. In this section we present a derivation of an observable equation for evolving the volume fraction.} 
For the purpose of capturing the interface while keeping pressure equilibrium at the interface, we consider the
case of an interface-only problem, following \citet{ShyueKM:98a}, in which velocity and pressure are constant in the
domain while density and material properties change across the interface. For simplicity, we present the one-dimensional
equations. \cref{eq:OEulerMass1,eq:OEulerMom1,eq:OEulerEn1} can be rewritten in the following
non-conservative form:
\begin{alignat}{2}
	 & \frac{{\partial \rho }}{{\partial t}}  &  & + \,\bar \rho \frac{{\partial u}}{{\partial x}} + \,\overline u \frac{{\partial \rho }}{{\partial x}} = 0,                                                                                                                                                                                                                                                                                                                                                 \\
	 & \rho \frac{{\partial u}}{{\partial t}} &  & + \overline {\rho u} \frac{{\partial u}}{{\partial x}} + \left( {\overline u  - u} \right)\frac{{\partial \rho u}}{{\partial x}}\, + \frac{{\partial p}}{{\partial x}} = 0,                                                                                                                                                                                                                                                                                \\
	 & \frac{{\partial \rho e}}{{\partial t}} &  & + \,\overline {\rho e} \frac{{\partial u}}{{\partial x}} + \bar u\frac{{\partial \rho e}}{{\partial x}} + \,\bar p\frac{{\partial u}}{{\partial x}} + \left( {\bar u - u} \right)\frac{{\partial p}}{{\partial x}} + \frac{1}{2}\left( {\overline {\rho {u^2}}  - 2u\overline {\rho u}   + \overline u \rho u} \right)\frac{{\partial u}}{{\partial x}}  {\rm{  + }}\frac{1}{2}\left( {{u^2} - u\bar u} \right)\frac{{\partial \rho u}}{{\partial x}} = 0,
\end{alignat}
where $e$ is the specific internal energy with $E = e + \frac{1}{2}{u^2}$. For the given interface-only problem,
the equations for the evolution of interface can be inferred to be:
\begin{alignat}{1}
	 & \frac{{\partial \rho }}{{\partial t}} + \,\overline u \frac{{\partial \rho }}{{\partial x}} = 0,   \\
	 & \frac{{\partial \rho e}}{{\partial t}} + \,\overline u \frac{{\partial \rho e}}{{\partial x}} = 0.
	\label{eqn:ObsIntEnergyForInterface}
\end{alignat}
By inserting the equation of state into \cref{eqn:ObsIntEnergyForInterface} and rearranging we have:
\begin{equation}
	\Gamma \left( {\frac{{\partial p}}{{\partial t}} + \overline u \frac{{\partial p}}{{\partial x}}} \right) + p\left( {\frac{{\partial \Gamma }}{{\partial t}} + \overline u \frac{{\partial \Gamma }}{{\partial x}}} \right) =0.
\end{equation}
With the pressure equilibrium requirement, the first parenthesis is zero. Since this equation needs to be
satisfied for any pressure, the term in the second parentheses should be zero. Writing second parenthesis using
\cref{eqn:mixturematerialprop}, the simplified equation is
\[\frac{{\partial z _1 }}{{\partial t}} + \overline u \frac{{\partial z _1 }}{{\partial x}} = 0,\]
which is the observable equation for capturing a material interface in one-dimension and in general form of the equation is:
\begin{equation}
	\frac{{\partial {z _1}}}{{\partial t}} + \overline {\bf{u}} \cdot \nabla {z _1} = 0.
	\label{Eqn:ObservableVolumeFraction}
\end{equation}
In order to validate and verify the applicability of
{\cref{eq:OEulerMass1,eq:OEulerMom1,eq:OEulerEn1}} and {\cref{Eqn:ObservableVolumeFraction}}
we perform several numerical experiments. In the next section, the numerical method used for this work is explained.
Here, we used ideal gas equation of state to derive the volume fraction equation as other authors have done in
their work \cite{KarniS:96a,AbgrallR:96a}. Using stiffened equation of state, as reported in \cite{BeigSA:15a},
which is suitable for liquids in addition to gases, the same equation for volume fraction will be achieved as
demonstrated in appendix A.

\subsection{Numerical Method}
\label{section-numerical-method}
As mentioned before, the observable regularization is done at the level of the PDE and not at the discretization
level. As a result, one can utilize any numerical technique in solving the observable equations. In the past we have
used many different numerical methods including high order finite difference compact Pad\'e in solving observable
Burgers' equation or 1D Euler equations in our group \cite{Mohseni:06l}. However, to show the robustness of the
regularization in the observable technique, here we used pseudo-spectral method, which provides a numerical technique
with no/minimal numerical dissipation as compared with other methods. Our objective is to show that a spectrally high
order technique can be used without any numerical dissipation.

The pseudo-spectral method is a standard method \mbox{\cite{FornbergB:98a,BoydJP:01a,JohnsonS:11a}}, which is used
here with no modifications. Basically, all spatial derivatives (e.g. $u_x$, $u_{xx}$, $u_{yy}$) are calculated in the
Fourier domain while all multiplications (e.g. $u \cdot \nabla u$) are done in the physical domain. An explicit method
is used for the time integration and a Helmholtz kernel is used for computing the observed quantities.

{\it \noindent Computing derivatives}.
Using pseudo-spectral for taking derivative and applying an averaging kernel such as Helmholtz kernel is very
	straightforward \mbox{\cite{FornbergB:98a,BoydJP:01a,JohnsonS:11a}}. Having a function $f$ on $N$ discrete points,
	${f'_n} = f'(nL/N)$, where $n$ is the index of the point and $L$ is the length of the domain. Assuming $N$ is even,
	the scaled wave number is defined as
\begin{equation}
	{k_S} = \left\{ {\begin{array}{*{20}{c}}
				{2\pi k/L}       & {0 \leqslant k < N/2} \\
				{2\pi (k - N)/L} & {N/2 \leqslant k < N}
			\end{array}} \right.
\end{equation}
and to take the first derivative of a function $f$,
	first, Fourier coefficients $F_k$ for $0 \leqslant k<N$ are calculated by taking FFT of $f$ using $f_n$ for $0
	\leqslant n < N$. Then, $F_k$ is multiplied by $i k_S$ for $k \ne N/2$ (where $i$ is the imaginary unit) and
	$F_{N/2}$ is multiplied by zero to obtain $F'_k$ \cite{JohnsonS:11a}. Finally, ${f'_n}$ is calculated by taking the
	inverse FFT of $F'_k$.
{\\ \it Buffer zone}.
Since a pseudo-spectral method is used here and a Fast Fourier Transform inherently assumes that the variable is
periodic, a buffer zone is added to the sides of the physical domain to satisfy the following conditions:
(I) periodicity of all the variables in the computational domain,
(II) smooth transition from the condition on the right to the left of the physical domain,
(III) prevent reflection from boundaries.
To achieve the second condition while satisfying the third condition we use a smooth $5^{th}$ order polynomial
as a weighting function to transition from the governing equation in the physical domain to a linear advection equation
with a damping term in the buffer zone.
The governing equation in the buffer zone is applied using similar ideas proposed by \citet{Freund:97a}. For the
weighting function, the following smooth step function is used:
\begin{equation}
	H(a,b,x) = \left\{ {\begin{array}{*{20}{l}}
				0                                        & {x \leqslant a} \\
				{6{\chi ^5} - 15{\chi ^4} + 10{\chi ^3}} & {a < x < b}     \\
				1                                        & {b \leqslant x}
			\end{array}} \right.
\end{equation}
where $\chi = \left( {x - a} \right)/\left( {b - a} \right)$.
For any governed quantity, here represented by $f$, the governing equation in the right and left buffer zones are
$\frac{{\partial f}}{{\partial t}} = - {U_R}\frac{{\partial f}}{{\partial x}} + \sigma \left( {{f_R} - f} \right)=RHS_R$
and $\frac{{\partial f}}{{\partial t}} = - {U_L}\frac{{\partial f}}{{\partial x}} + \sigma \left( {{f_L} - f}
	\right)=RHS_L$, respectively. The first term in these equations is a linear advection with a prescribed constant
velocity $U_R$ and $U_L$ which is set equal to the initial horizontal velocity at the right and left edge of physical
domain, respectively. The second term in the buffer zone equations is a sponge term that brings the quantity $f$ towards
the prescribed values $f_R$ and $f_L$, which are again set based on initial value of $f$ at the right and left edge of
physical domain. $\sigma$ is a damping coefficient which is here defined to be $1/\Delta x$. In order to find the right
hand side of the equation in the buffer zones, the $x$ location of the right boundary is mapped linearly to $\xi \in
	[0,0.5)$ and the $x$ location of the left buffer zone mapped to $\xi \in [0.5,1)$. Then the right hand side is defined
as
\begin{equation}
	RH{S_{\mathit{Buffer}}} = H\left( {a,b,\xi } \right)RH{S_{Physical}} + \left[ {1 - H\left( {a,b,\xi } \right)} \right]\left[ {H\left( {c,d,\xi } \right)RH{S_L} + \left( {1 - H\left( {c,d,\xi } \right)} \right)RH{S_R}} \right],
	\label{eqn:bufferZoneRHS}
\end{equation}
where $RHS_{Physical}$ is the right hand side of the governing equation in the physical domain. $a$, $b$, $c$, and $d$
are 0.1, 0.5, 0.5, and 0.9, respectively, which specify the start and end of the smooth transitions. The values of these
constants will be mentioned for each problem in the results section. Note that the buffer zone setup presented here is
not necessarily the best choice, but it satisfies the three major requirements mentioned above for the problems solved
in this manuscript. 
{\\ \it Computation of observed quantities}. It is demonstrated for the regularized Burgers'
equation that using different averaging kernels, including Gaussian and Helmholtz, has negligible effect on the large
scale features, features above the observability scale $\alpha$, \cite{Mohseni:08w,Mohseni:09h}. While any kernel
can be used to implement the method proposed in this work, since applying Helmholtz kernel is equivalent to simple
algebraic operations in the Fourier space, without loss of generality (see \cite{Mohseni:09h}) and for the sake of
simplicity, we use this kernel as the averaging kernel defined by,
\begin{equation}
	{g^\alpha }(x) = {C_\alpha }{e^{\left( { - \left| x \right|/\alpha } \right)}},
\end{equation}
where $C_\alpha$ is a normalization constant. Alternatively, the Helmholtz kernel can also be applied by solving the
Helmholtz equation:
\begin{equation}
	f = \overline f  - {\alpha ^2}{\nabla ^2}\overline f.
	\label{Eqn:Helmholtz}
\end{equation}
Computing $\overline f$ using a Helmholtz operator, \mbox{\Cref{Eqn:Helmholtz}}, is easily done by multiplying $F_k$
by $1/(1+\alpha ^2 k_S^2)$ and taking an inverse FFT.
{\\ \it Time integration and dealiasing}. For the time integration, a $3^{rd}$ order TVD Runge-Kutta method is
used \cite{ShuCW:98a}. For dealiasing, a low pass exponential filter, $G(k) = \exp ( { - 36{{\left| k/k_{\max }
\right|}^{36}}} )$, is used \cite{HouTY:07a}. In a similar manner to applying Helmholtz kernel, this low pass filter
is computed by multiplying $F_k$ by $G(k_S)$ and computing the inverse FFT of the result. The dealiasing filter is
applied to all conservative variables in addition to the volume fraction after each Runge-Kutta step. For all initial
conditions double filtering is used as explained in \cite{Mohseni:10k}.
\section{Results and discussion}
Here, we present several 1D and 2D test cases to show the performance of the proposed method compared to other available
methods, experiments and exact solutions.
\subsection{One-dimensional advection of isolated interface}
First, an isolated interface is examined for a one-dimensional domain with periodic boundary condition and the initial
condition defined by
\begin{equation}
	\begin{gathered}
		{\left( {\rho ,u,p,\gamma } \right)^T}_{x \leqslant 0} = (1,0.5,1/1.4,1.4)^T, \hfill \\
		{\left( {\rho ,u,p,\gamma } \right)^T}_{x > 0} = (10,0.5,1/1.4,1.2)^T. \hfill \\
	\end{gathered}
\end{equation}

\begin{figure}[htbp]
	\begin{center}
		\includegraphics[width=5.8 in]{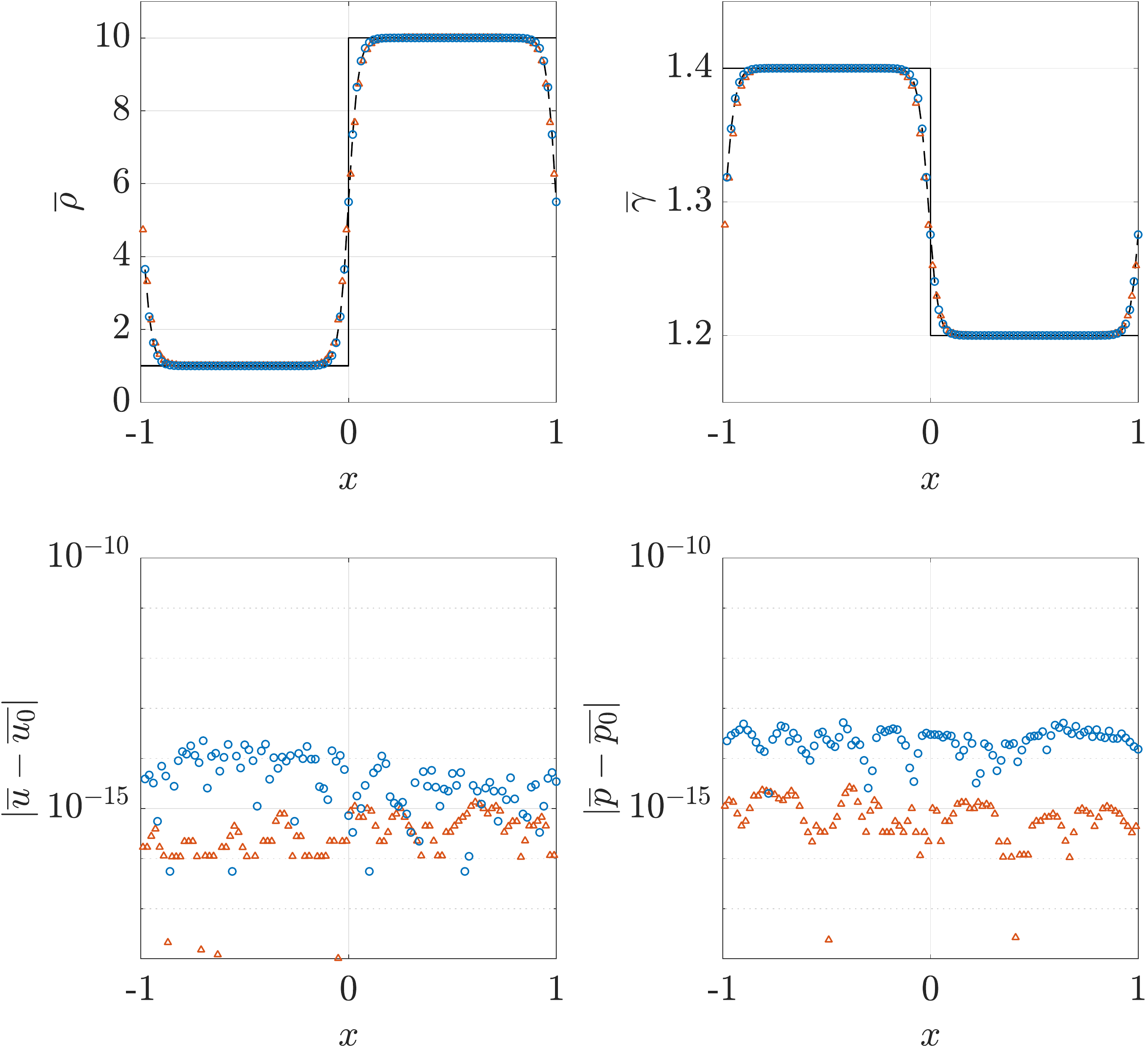}
		\caption{Solution of isolated interface problem using observable two-phase Euler equation (blue open circles)
		compared to the exact solution of two-phase Euler equation (black solid lines) and solution of two-phase Euler
		problem using a WENO5 finite volume method,  \citet{Colonius:06a} (red open triangles). The initial
		condition for the observable equations is shown as a black dashed line. Mixture density, $\overline \rho$ (top
		left), ratio of specific heat $\gamma$ (top right), magnitude of difference between final and initial velocity
		(bottom left), and magnitude of difference between final and initial pressure (bottom right) are shown. }
		\label{Fig:IsolatedInterface}
	\end{center}
\end{figure}

Since the domain is periodic and velocity is constant across the doma in, after one time period, at $t=4$, all
the flow variables should be exactly equal to initial condition as shown in \cref{Fig:IsolatedInterface}. Observed
density and ratio of heat capacities are shown using blue open circles which are in good agreement with the
exact solution of the Euler equation (black solid lines). In addition the results of a WENO5 finite volume method
calculated by \citet{Colonius:06a} (orange open triangles) is included. The magnitude of differences between the final
and initial observed velocity and pressure are demonstrated to be zero to the order of machine precision, as
expected. This shows that our method preserves the pressure equilibrium at material interfaces. The grid resolution is
the same as \cite{Colonius:06a}, $\Delta x = 0.01$, and non-dimensional observability limit, $\alpha / \Delta x$, is
set to $0.75$.
\subsection{One-dimensional interaction of a shock wave with a Helium bubble}
\label{subsec:shock_helium_bubble}
\begin{figure}[htbp]
	\begin{center}
		\includegraphics[width=5.8 in]{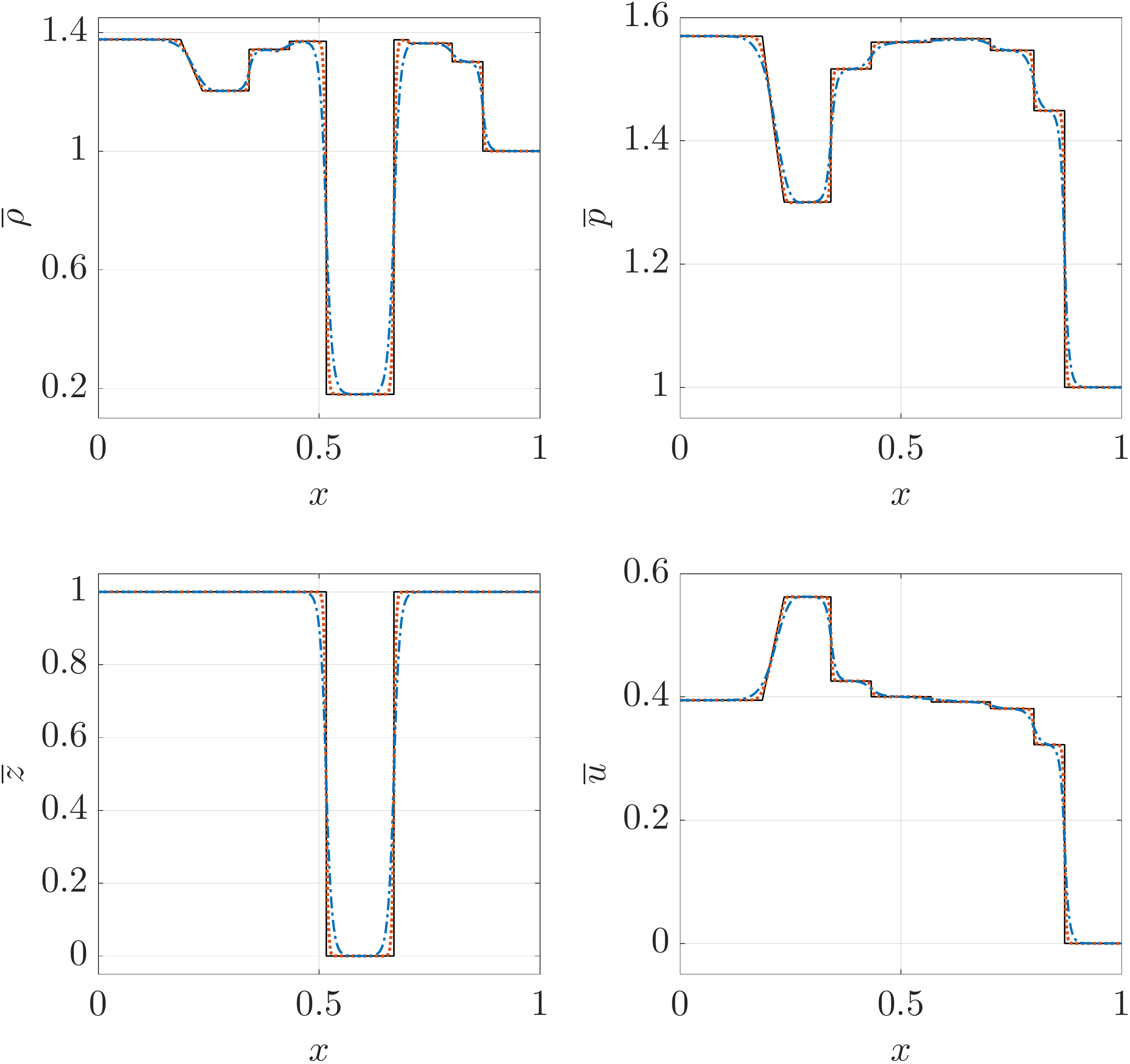}
		\caption{Plot of observed density, pressure, volume fraction, and velocity for a shock-Helium bubble interaction
			at $t=0.35$ with initial condition given by \cref{Eqn:Shock-HeBubble-Init}. Solid line is the exact Euler
			solution. Red dotted line and blue dash-dot line are simulations with 1024 grid points using
			$\alpha /\Delta x$ of 1.25 and 3.75, respectively. }
		\label{Fig:1DShock-He-Bubble}
	\end{center}
\end{figure}

Next a shock-bubble interaction, proposed by \citet{KarniS:96a}, is studied. In this problem a Mach 1.22 shock
wave traveling in air interacts with a Helium bubble. This interaction leads to different wave structures that show the
performance of the method in handling all possible waves in an inviscid compressible flow, rarefaction wave, shock wave,
and contact discontnuity. The initial condition for this problem is
\begin{equation}\left( {{\rho _1}z ,{\rho _2}(1 - z ),u,p,z } \right) = \left\{ {\begin{array}{*{20}{l}}
				{\left( {{\rm{1.3764}},0,{\rm{0.3947}},{\rm{1.5698}},1} \right)} & {0 \le x < 0.25,}   \\
				{\left( {1,0,0,1,1} \right)}                                     & {0.25 \le x < 0.4,} \\
				{\left( {0,0.138,0,1,0} \right)}                                 & {0.4 \le x < 0.6,}  \\
				{\left( {1,0,0,1,1} \right)}                                     & {0.6 \le x < 1.}
			\end{array}} \right.
	\label{Eqn:Shock-HeBubble-Init}
\end{equation}
The ratio of specific heat, $\gamma$, is set to 1.67 and 1.4, inside and outside the bubble,
respectively. \Cref{Fig:1DShock-He-Bubble} shows the results for this problem at $t=0.35$. The solid line is the exact
Euler solution. The red dotted line and blue dash-dot line are simulations using 1024 grid points with $\alpha/\Delta x$
equal to 1.25 and 3.75, respectively. 

When the incident shock hits the interface, it reflects a left going rarefaction wave while a shock wave passes
into the bubble. This shock hits the other end of the bubble and creates a transmitted shock and a reflected
shock. These interactions consequently result in two left going and three right going shock waves outside the
bubble and a very weak (almost invisible) left going shock wave inside the bubble until $t=0.35$. As shown in
the figure, for this problem the method can regularize the equation with a non-dimensional observability limit as small
as 1.25. Obviously, as $\alpha/\Delta x$ increases, the equations become more regularized but at the same time less
details will be captured for any quantities of interest.
\subsection{One dimensional interaction of a strong shock wave with an air bubble}
As stated by \citet{Colonius:14a}, the interaction of a shock with a material interface can challenge numerical
methods that do not solve the equations in conservative form. The main known artifact is likely error in the resulting
wave speeds, which deteriorate as time passes and the error in mass conservation increases. To investigate the
performance of our method in this regard, the interaction of a strong shock wave with a material interface is studied
here.

\begin{figure}[htbp]
	\begin{center}
		\includegraphics[width=\textwidth]{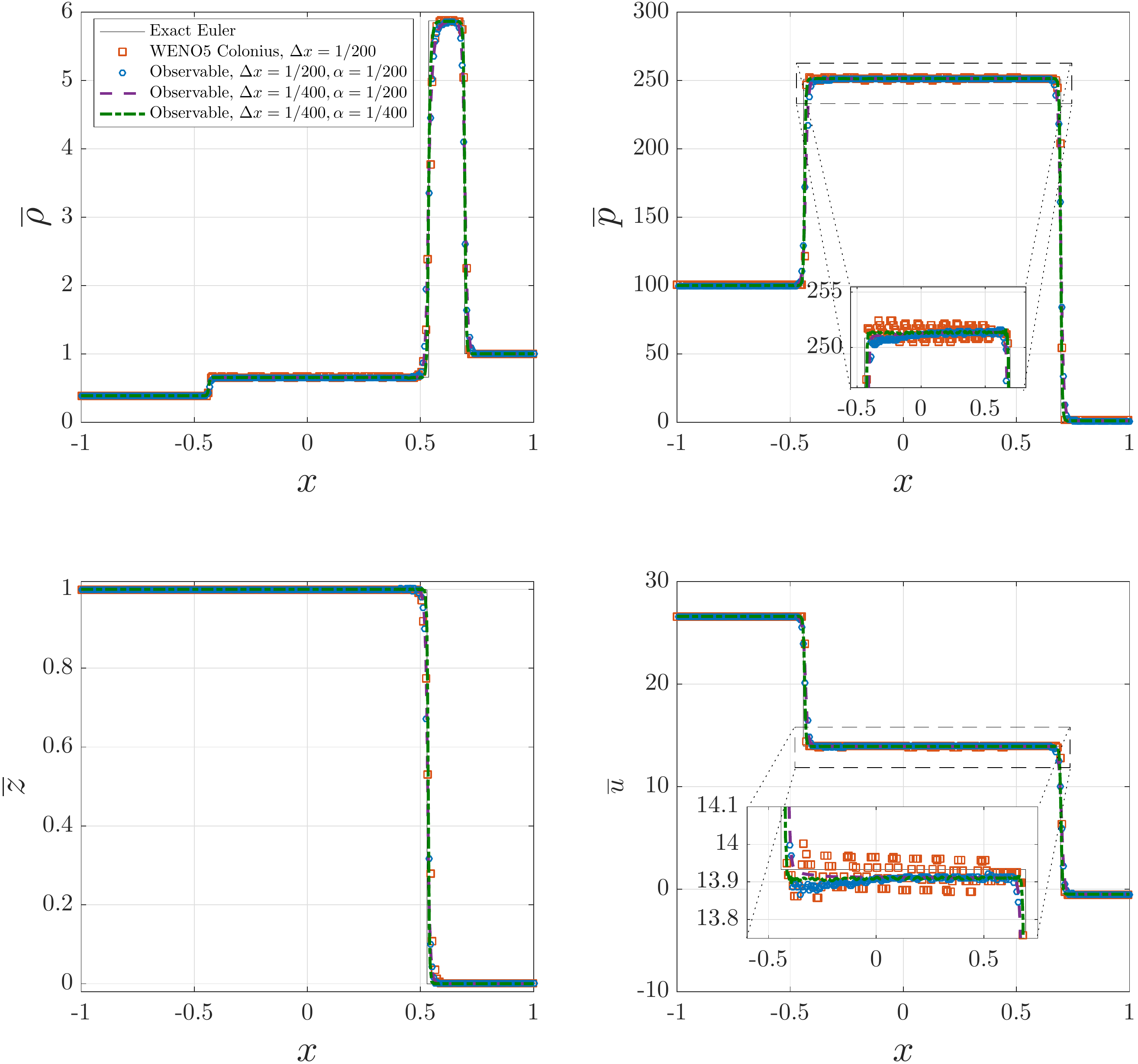} 
		\caption{Plot of observed density, pressure, volume fraction, and velocity for a shock-bubble interaction at
			$t=0.07$ with initial condition given by \cref{Eqn:Shock-AirBubble-Init}. Blue open circles show the result
			with the present method, red open squares show the results calculated using a WENO5 finite volume method
			presented by \citet{Colonius:14a}, and solid lines show the exact Euler solution. The insets in the pressure
			and velocity plots show an enlarged portion of the results and demonstrate that the observable method avoids
			small amplitude non-physical oscillations that appears in the WENO5 results. \addtxt{As the ratio of
			$\alpha/\Delta x$ increases the non-physical oscillations are reduced as shown using a simulation with a
			higher resolution and the same $\alpha$. Keeping $\alpha/\Delta x$ constant and increasing resolution on the
			other hand makes the shock and interface sharper but still we can see very small high wave mode
			oscillations, dash-dot green line.}}
		\label{Fig:1DShock-Air-Bubble}
	\end{center}
\end{figure}

This problem is originally proposed by \citet{LiuT:03a}, and then a modified form of it, as used here, is
solved by \citet{JohnsenE:07a} and \citet{Colonius:14a}. In this problem a Mach 8.96 shockwave moving in Helium
interacts with an air bubble. Initially, the unshocked region has a small velocity towards the shock. The same
resolution as \cite{Colonius:14a} is used, $\Delta x = 0.005$, with a non-dimensional observability limit $\alpha /
	\Delta x = 1$.  A buffer zone of length 2 is used on both sides of the physical domain. The initial condition is
\begin{equation}
	\left( {{\rho _1}z ,{\rho _2}(1 - z ),u,p,z } \right) = \left\{ {\begin{array}{*{20}{l}}
				{\left( {0.386,0,26.59,100,1} \right)} & { - 1 \le x <  - 0.8,}   \\
				{\left( {0.1,0, - 0.5,1,1} \right)}    & { - 0.8 \le x <  - 0.2,} \\
				{\left( {0,1, - 0.5,1,0} \right)}      & { - 0.2 \le x < 1.}
			\end{array}} \right.
	\label{Eqn:Shock-AirBubble-Init}
\end{equation}
\Cref{Fig:1DShock-Air-Bubble} shows good agreement between our results and the exact solution and demonstrate
that all the wave velocities are captured correctly. Since we are using pseudo-spectral method to calculate derivatives,
any sharp changes need to be taken place over several grid points in order to avoid the Gibbs phenomena. As a result of
the strong shock, using a non-dimensional observability limit of less than unity creates spurious oscillations in the
result, which would be expected as it means the computation is under-observed. Using the Helmholtz kernel, the
observability limit imposes the shock thickness of about $4.6\alpha$ (which here translates to about 5 cells or 6 grid
points), as shown by \citet{Mohseni:10k}.  While, the WENO5 computation of  \citet{Colonius:14a} is demonstrating sharper
interface and shockwave results using the same resolution, it also creates small amplitude non-physical oscillations
noticeable in the pressure and velocity plots for $x \in (-0.4,0.5)$ while this spurious oscillation is not seen in
observable results.
\subsection{Two-dimensional interaction of a shock wave with a R22 bubble}
\begin{figure}[htbp]
	\begin{center}
		\setlength{\unitlength}{0.165\linewidth}
		\includegraphics[width=5.5\unitlength]{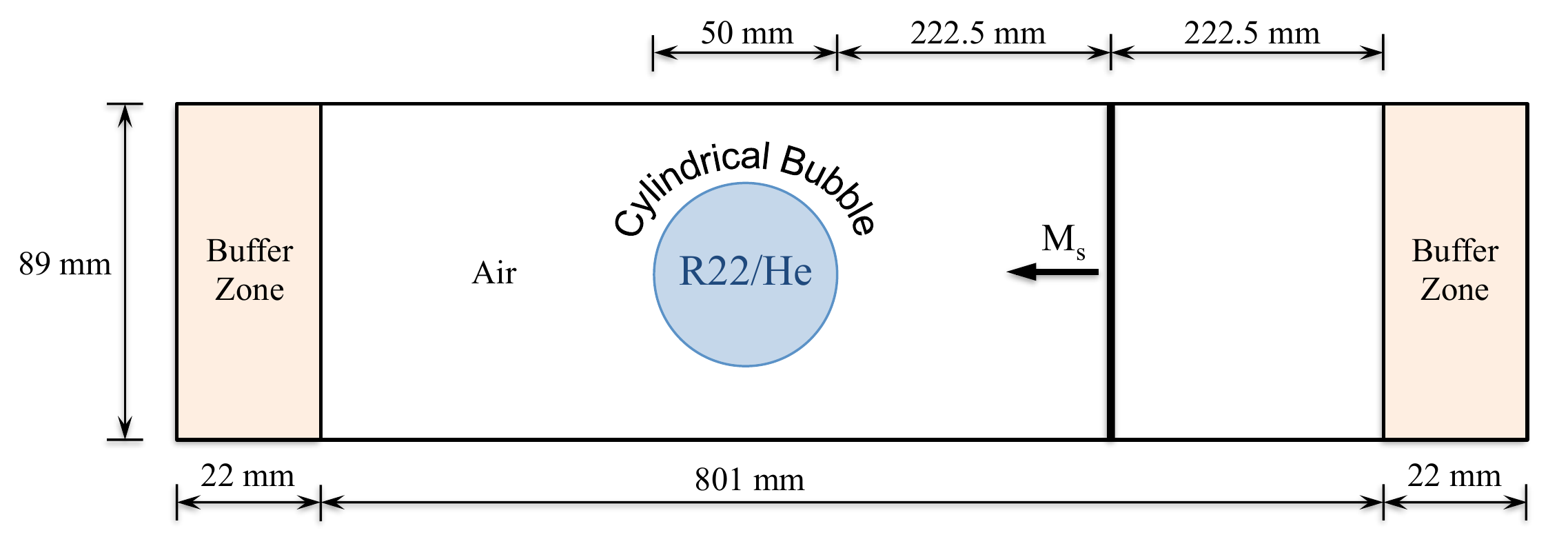}
		\caption{Schematic of the shock-bubble problem. Buffer zones are used at the right and left boundary to make the
			numerical domain periodic for the purpose of pseudo-spectral calculations. In addition, equations in the
			buffer zone are set to behave like a non-reflective boundary. A periodic boundary condition is assumed at top
			and bottom boundaries.}
		\label{Fig:2DShock_R22_Bubble_Schematic}
	\end{center}
\end{figure}
In this section our method is validated with available experimental results. This problem is a two-dimensional
counterpart of the one that is solved in \cref{subsec:shock_helium_bubble}; a Mach 1.22 shockwave in air interacts with
a cylindrical R22 bubble with $\gamma = 1.249$ and $\rho_{R22} = 3.712\, kg/m^3$. This problem was first studied
experimentally by \citet{SturtevantB:87a} and soon became a benchmark for numerical methods,
\cite{TheofanousT:06a,BanksJ:07a,KreeftJ:10a,KokhS:10a,RanjanD:08a,ChertockA:08a,MaoD:13a}, since it has
complex wave interactions and also contains a Richtmyer-Meshkov instability that poses a severe challenge on
computational methods. The current simulation is done using a grid resolution of $\Delta x = 222\,\mu m$ and a
non-dimensional observability limit $\alpha/\Delta x = 1$. For plotting numerical results, the visualization technique
introduced by \citet{KarniS:96a} is used here. \Cref{Fig:2DShock_R22_Bubble_Schematic} shows the schematic of the
problem setup. At the left and right side of the domain, a buffer zone is used to make the numerical domain periodic
(for the purpose of pseudo-spectral computations). Additionally, the buffer zone damps the outgoing waves and minimizes
reflection from boundary. The top and bottom boundaries assume a periodic boundary condition which, with the symmetry of
the problem, act like a reflective boundary condition. In order to initialize the curved interface on a rectangular grid
and minimize the stairs-like initialization of the interface, which could lead to additional waves and interface
instability in compressible flow, here a $tanh$ function is used for the definition of initial volume fraction,
$z(x,y) = 0.5+0.5*tanh((D(x,y)-r_{bubble})/\delta)$, where $r_{bubble}$ is the bubble radius and $D(x,y)$ is the
distance from the center of the bubble equal to $\sqrt{(x-x_c)^2+(y-y_c)^2}$, and $\delta$ is approximately one-third of
the interface thickness which is set to 1.5 here.

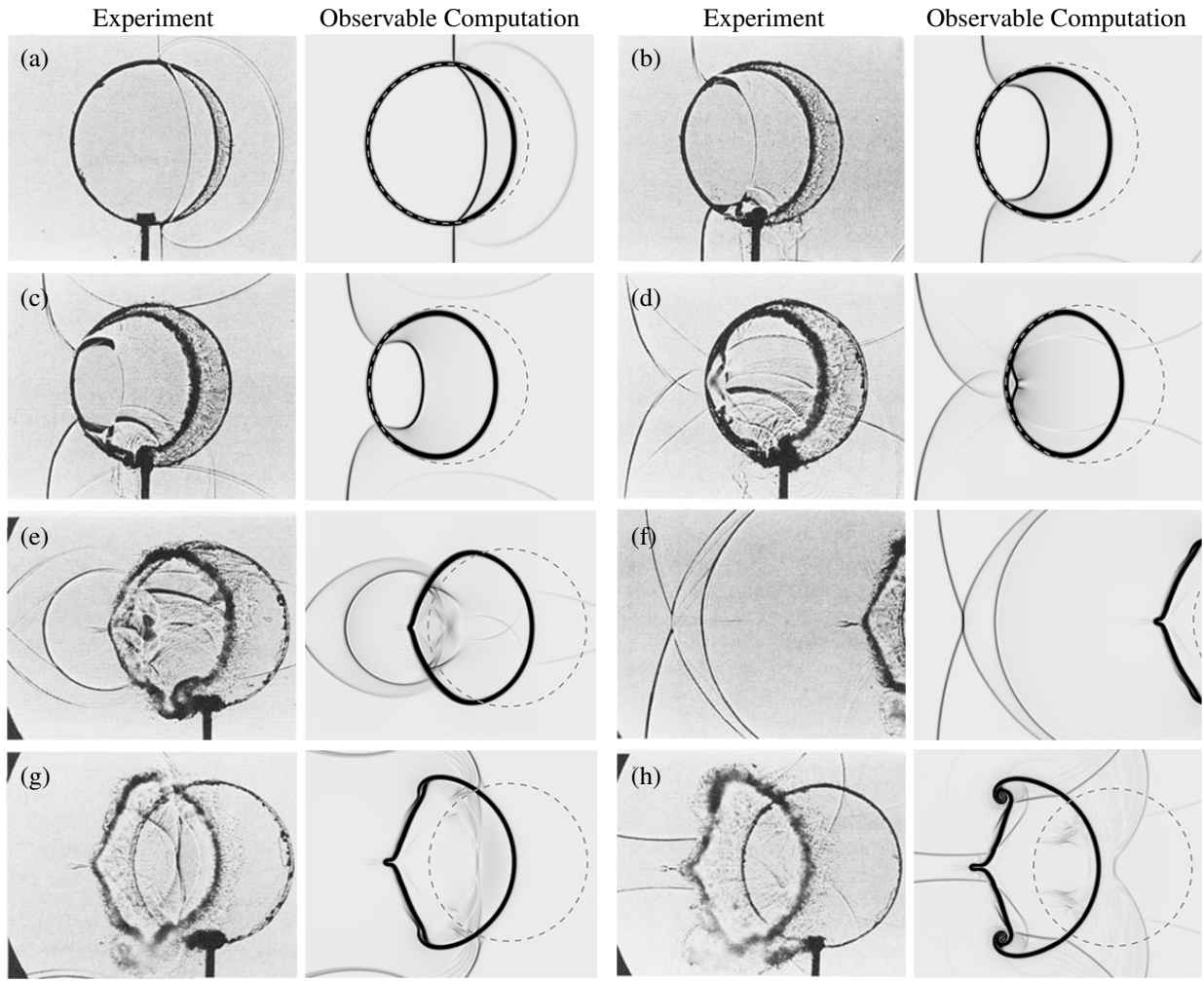
\begin{figure}[htbp]
	\begin{center}
		\setlength{\unitlength}{0.165\linewidth}
		\begin{picture}(6,4.8)
			\put(0,0){\includegraphics[width=6\unitlength]{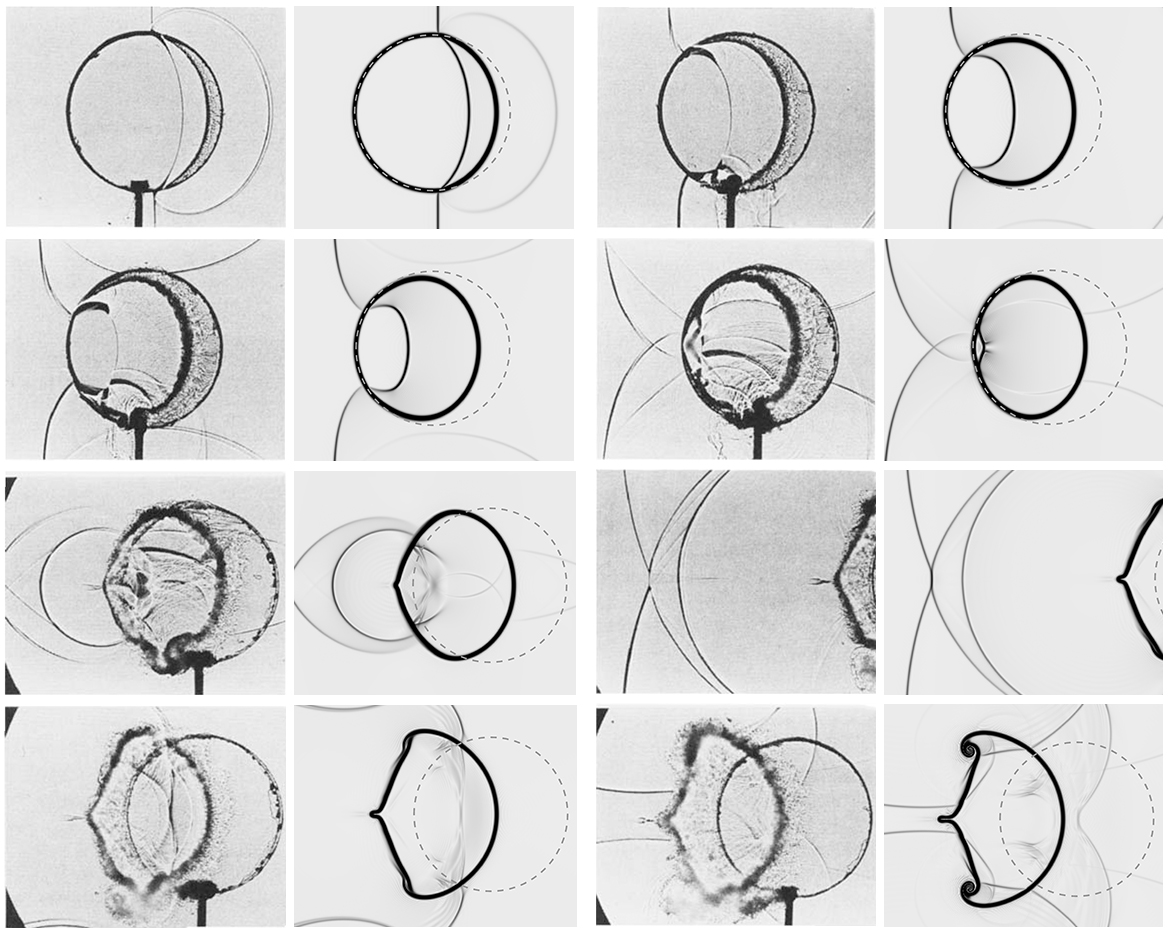}}
			\put(0.45,4.80){Experiment}
			\put(1.58,4.80){Observable Computation}
			\put(3.49,4.80){Experiment}
			\put(4.6,4.80){Observable Computation}
			\put(0.09,4.60){(a)}
			\put(3.13,4.60){(b)}
			\put(0.09,3.40){(c)}
			\put(3.13,3.40){(d)}
			\put(0.09,2.20){(e)}
			\put(3.13,2.20){(f)}
			\put(0.09,1.00){(g)}
			\put(3.13,1.00){(h)}
		\end{picture}
		\caption{Schlieren snapshots of the interaction of a planar $M_s=1.22$ shock in air with a cylindrical
			Refrigerant 22 (R22) bubble. The right frame of each subfigure shows the numerical schlieren of observable
			simulations while the left frames show the experimental snapshots from \citet{SturtevantB:87a}. The snapshots
			are taken at times (a) 55 $\mu s$, (b) 115 $\mu s$, (c) 135 $\mu s$, (d) 187 $\mu s$, (e) 247 $\mu s$, (f) 318
			$\mu s$, (g) 342 $\mu s$, and (h) 417 $\mu s$. For the observable simulation, a resolution of
			$\Delta x = 222\, \mu m$ and a non-dimensional observability limit of $\alpha/\Delta x = 1$ are used. The
			dashed line in numerical schlieren and solid circular line in experimental ones are the initial position of
			the bubble.}
		\label{Fig:2DShock_R22_Bubble}
	\end{center}
\end{figure}

\begin{figure}[htbp]
	\begin{center}
		\setlength{\unitlength}{0.135\linewidth}
		\begin{picture}(6,4.8)
			\put(0,0){\includegraphics[width=6\unitlength]{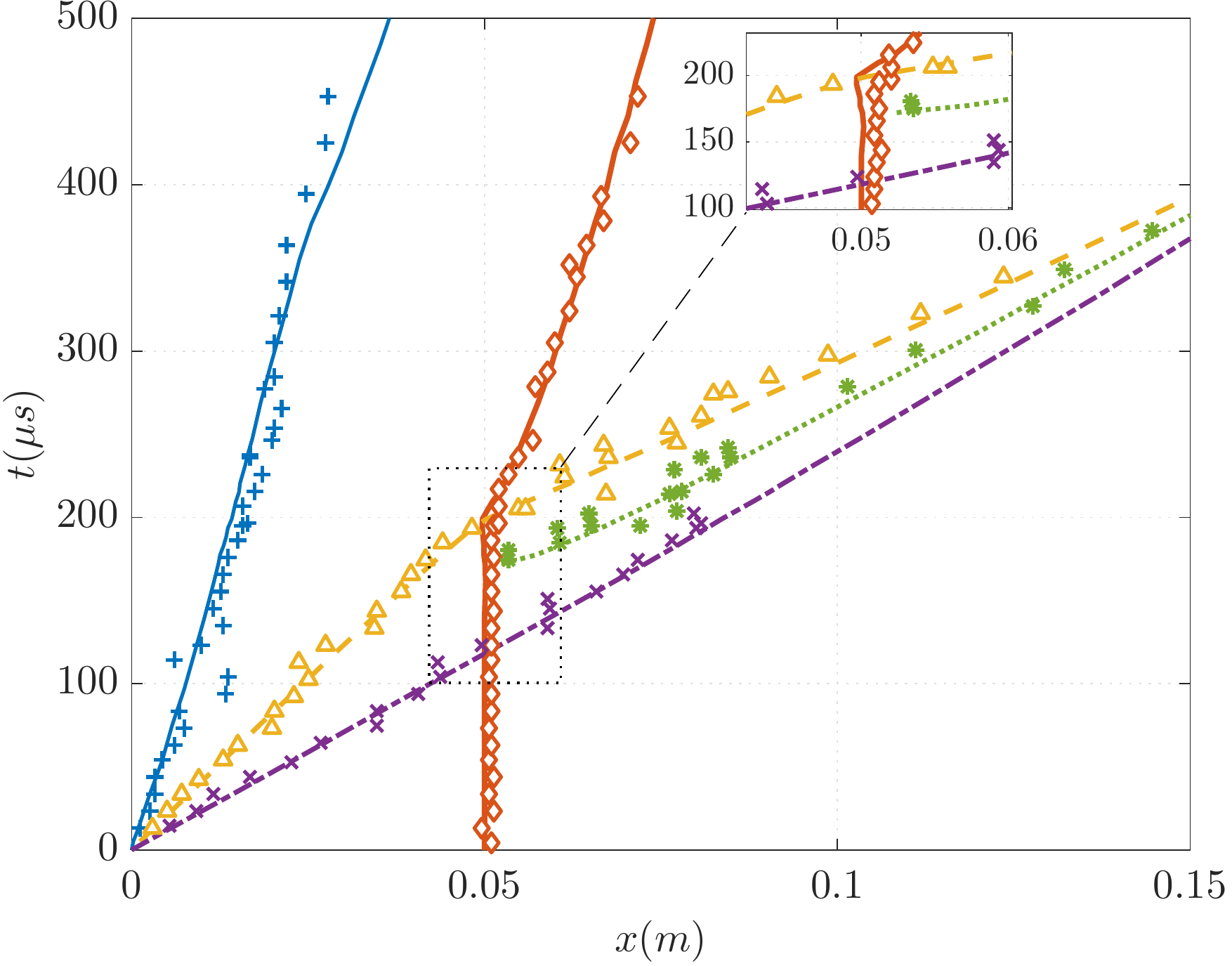}} 
			\put(4.2,0.65){{\includegraphics[width=1.5\unitlength]{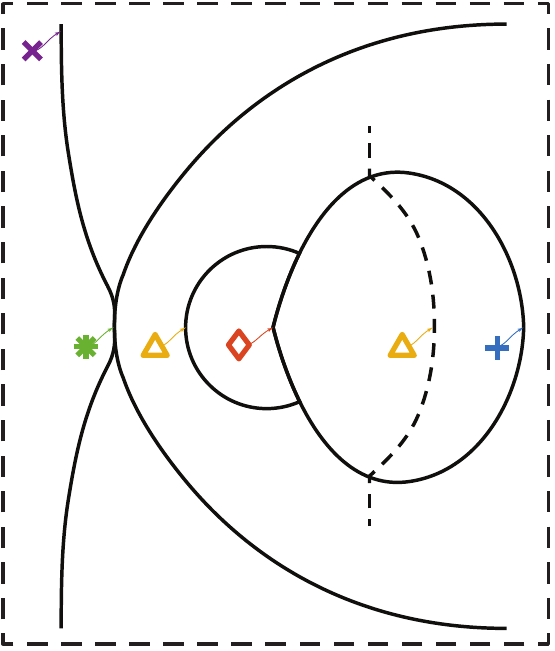}}}
		\end{picture}
		\caption{Various wave and interface locations during a Mach 1.22 shock-R22 cylindrical bubble interaction in
			addition to a small schematic of this interaction. The figure compares experimental results from Haas \&
			Sturtevant 1987 \cite{SturtevantB:87a} (symbols) with observable results (lines). \textcolor{Mpurple}{\bf
				\footnotesize $\pmb\times$}: incident shock, \textcolor{Mblue}{\bf \tiny $\pmb{+}$}: upstream edge of the
			bubble, \textcolor{Mred}{\bf \scriptsize $\pmb\Diamond$}: downstream edge of the bubble,
			\textcolor{Myellow}{\bf \scriptsize $\pmb\vartriangle$}: refracted shock, \textcolor{Mgreen}{\bf \scriptsize
				$\pmb\convolution$}: location of where the original diffracted shocks meet as shown in the schematic.}
		\label{Fig:shock_R22_bubble_wave_location_comparison}
	\end{center}
\end{figure}

\Cref{Fig:2DShock_R22_Bubble} shows experimental schlieren images of shock-cylindrical bubble interaction from
\citet{SturtevantB:87a} in addition to numerical schlieren from our results. The incident shock hits the interface at
$t= 0\,\mu s$; a shock wave reflects from the interface and another one refracts into bubble. Since the speed of sound
is lower inside the bubble, the refracted shock moves with a lower speed compared to the incident shock outside the
bubble, and as a result the shock converges towards the end of bubble; see snapshots (a), (b), and (c). The shock
initially reflected from the interface reflects from top and bottom wall, as shown in snapshot (c). The incident shock
diffracted around the bubble arrives at the downstream edge of bubble faster than the refracted shock (snapshot
d). Consequently, the pressure outside the bubble becomes larger than the pressure of unshocked gas inside the bubble
for a short period of time until the refracted shock reaches the downstream edge of the bubble. During this time period
the downstream edge starts deflecting towards the upstream edge; this can be noticed by looking closely at red solid
line in \cref{Fig:shock_R22_bubble_wave_location_comparison} between $t=161\,\mu s$ and $t=194\,\mu s$. This deflection
is very small and when the refracted shock, which is focused at the downstream edge, hits the interface it pushes back
the interface towards the downstream and creates a jet at that point, see snapshots (e) and (f). It is interesting to
note that depending on bubble size, shock strength, and the material property of the bubble and surroundings, a reverse
jet may run towards upstream, and even pierce the front edge of bubble in spherical bubble case, instead of a jet
towards downstream; for such examples see \cite{RanjanD:08a} and \cite{GiordanoJ:06a}. If the initial
defracted shock has enough time to make a dent into the downstream of the bubble before the refracted shock reaches
downstream edge of the interface, the high pressure, created as the refracted shock transmits out of the bubble,
pushes the trapped air in the dented area towards upstream and creates a reverse jet as opposed to what is seen here.
After the refracted shock is focused at the downstream edge of the bubble, it transmits in an almost circular shape,
trailing the defracted wave, snapshots (e) and (f). Interaction of the shock with the interface creates rollup and
instability of the interface which become visible in snashots (g) and (h).

The results in \cref{Fig:2DShock_R22_Bubble} show good agreement between experimental and simulation snapshots. All the
main waves and interfaces are captured correctly. To show the accuracy of the results with respect to the experiment,
several wave and interface locations over time are extracted from the simulation and compared in
\cref{Fig:shock_R22_bubble_wave_location_comparison} with what is reported in experiment \cite{SturtevantB:87a}. In this
figure the lines show our results while symbols show the result from the experiment. A small schematic of this shock
interaction problem shows the location of different waves and interfaces where data are extracted.  As shown in
\cref{Fig:shock_R22_bubble_wave_location_comparison}, the observable results are in good agreement with the reported
experimental results. A quantitative comparison of different wave and interface speeds is shown in \cref{Table:Shock-R22-Bubble-Velocities}, including experimental results, numerical results from \citet{KarniS:96a} and \citet{ShyueKM:06a},
in addition to the observable results of the current simulation. The reported observable wave and interface speeds are
calculated by linear least square fitting the data shown in \cref{Fig:shock_R22_bubble_wave_location_comparison} for the
time intervals mentioned in the caption of the table. As shown in the table, all the reported speeds from the observable
method are in good agreement with experimental results and are well within the 10\% error reported for the experimental
results. Since the resolution of our simulation is at least about twice as coarse as the other reported simulations,
while producing similar or better results (e.g. $V_{di}$) compared to the experiment, we estimate that the current
method reduces the computational cost by about an order of magnitude in two-dimensions (2 times less grid points in
each direction and twice larger time step size based on CFL criterion). Obviously, the savings will be much higher in
three-dimensional computations if the current method is used.

\begin{table}[htbp]
	\centering
	\caption{Comparison of wave and interface velocities in a Mach 1.22 shock-R22 bubble interaction from experiment
		\cite{SturtevantB:87a} with different numerical results from literature and observable results. The grid resolution
		for each computational result is shown in the parenthesis. $V_s$: incident shock velocity, $V_R$: refracted shock
		velocity, $V_T$: transmitted shock velocity, $V_{ui}$: initial velocity of upstream edge of bubble, $V_{\mathit{uf}}$:
		final velocity of upstream edge of bubble, and $V_{di}$: initial velocity of downstream edge of bubble. Observable
		results are calculated by a linear least square fitting during the following time periods,
		$0 \,\mu s \leqslant t_{V_s} \leqslant 172 \,\mu s$, $5 \,\mu s \leqslant t_{V_R} \leqslant 172 \,\mu s$,
		$199 \,\mu s \leqslant t_{V_T} \leqslant 290 \,\mu s$, $0 \,\mu s \leqslant t_{V_{ui}} \leqslant 172 \,\mu s$,
		$376 \,\mu s \leqslant t_{V_{\mathit{uf}}} \leqslant 506 \,\mu s$, and
		$199 \,\mu s \leqslant t_{V_{di}} \leqslant 269 \,\mu s$.}
	\vspace{2mm}
	\begin{tabular}{@{}lcccccc@{}}
		\toprule
		                                                            & $V_s$ & $V_R$ & $V_T$ & $V_{\mathit{ui}}$ & $V_{\mathit{uf}}$ & $V_{di}$ \\ \midrule
		Experiment \cite{SturtevantB:87a}                           & 415   & 240   & 540   & 73                & 90                & 78       \\
		Quirk and Karni \cite{KarniS:96a} ($\Delta x=56\,\mu m$)    & 420   & 254   & 540   & 74                & 90                & 116      \\
		Shyue \cite{ShyueKM:06a} (tracking, $\Delta x=125\,\mu m$)  & 411   & 243   & 538   & 64                & 87                & 82       \\
		Shyue \cite{ShyueKM:06a} (capturing, $\Delta x=125\,\mu m$) & 411   & 244   & 534   & 65                & 86                & 98       \\
		Observable (this paper, $\Delta x=222\,\mu m$)              & 424   & 247   & 530   & 73                & 88                & 78       \\ \midrule
		\% Error: Quirk and Karni \cite{KarniS:96a}                 & 1.2   & 5.8   & 0.0   & 1.4               & 0.0               & 48.7     \\
		\% Error: Shyue \cite{ShyueKM:06a} (tracking)               & 1.0   & 1.3   & 0.4   & 12.3              & 3.3               & 5.1      \\
		\% Error: Shyue \cite{ShyueKM:06a} (capturing)              & 1.0   & 1.7   & 1.1   & 11.0              & 4.4               & 25.6     \\
		\% Error: Observable                                        & 2.2   & 2.9   & 1.9   & 0.0               & 2.2               & 0.0      \\\bottomrule
	\end{tabular}
	\label{Table:Shock-R22-Bubble-Velocities}
\end{table}

\subsection{Two-dimensional interaction of a shock wave with a Helium bubble}
Haas \& Sturtevant \cite{SturtevantB:87a} also conducted their experiment for a Mach 1.22 shock wave in the air
interacting with a cylindrical Helium bubble contaminated with 28\% air by mass.  In this case, the material properties
inside the bubble are $\gamma = 1.648$ and $\rho_{He+28\% Air} = 0.214\, kg/m^3$. The problem setup is exactly the same
as the shock-R22 bubble problem explained in the previous section. Since the speed of sound is now higher inside the
bubble compared to its surrounding, the refracted shock diverges as opposed to the converging case for the R22 bubble.
Because of a higher speed of sound inside the bubble and a higher variation of flow variables across the interface and
shocks compared to the R22 bubble case, this case is numerically more challenging.

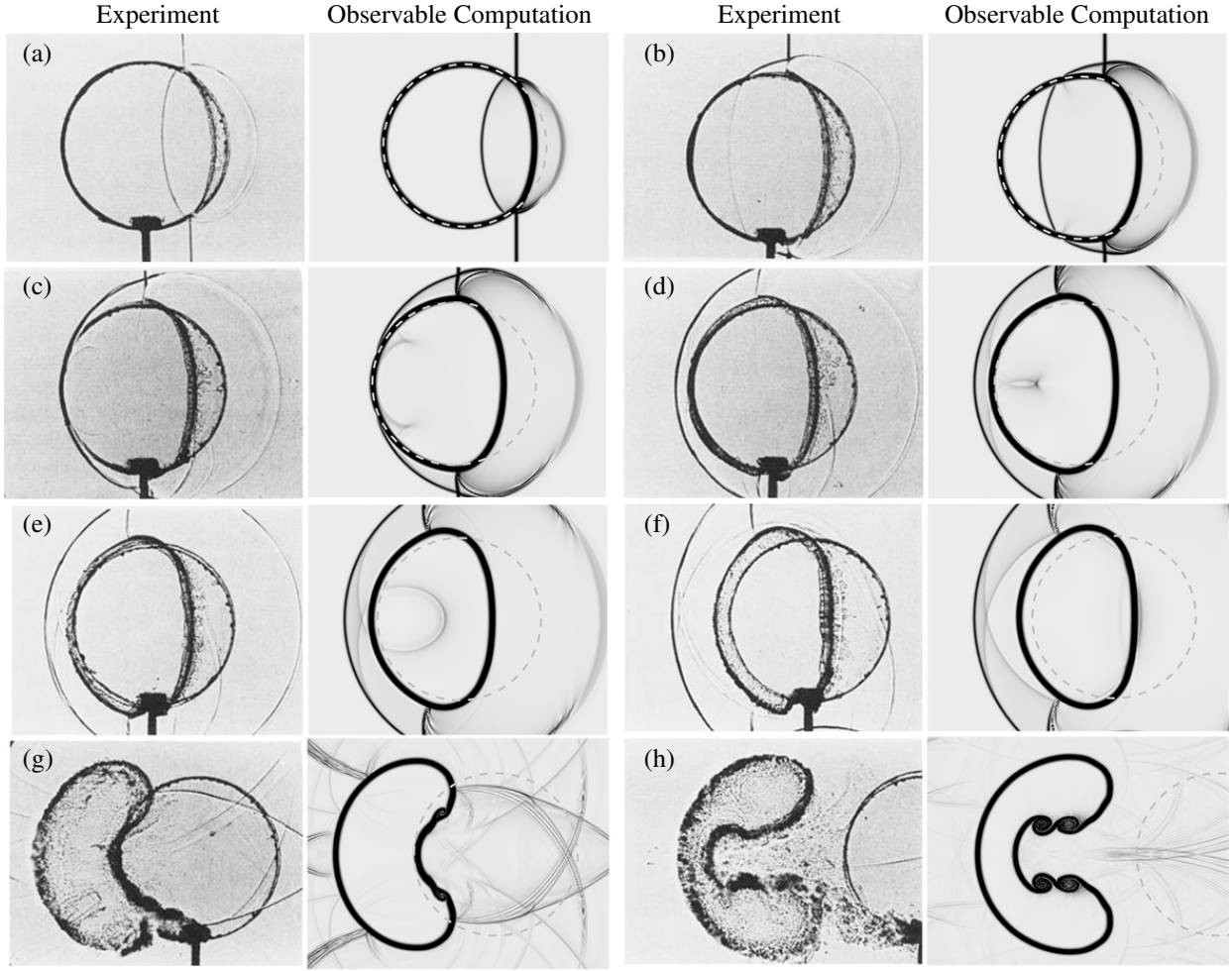
\begin{figure}[htbp]
	\begin{center}
		\setlength{\unitlength}{0.165\linewidth}
		\begin{picture}(6,4.8)
			\put(0.45,4.69){Experiment}
			\put(1.58,4.69){Observable Computation}
			\put(3.49,4.69){Experiment}
			\put(4.6,4.69){Observable Computation}
			\put(0,0.0){\includegraphics[width=6\unitlength]{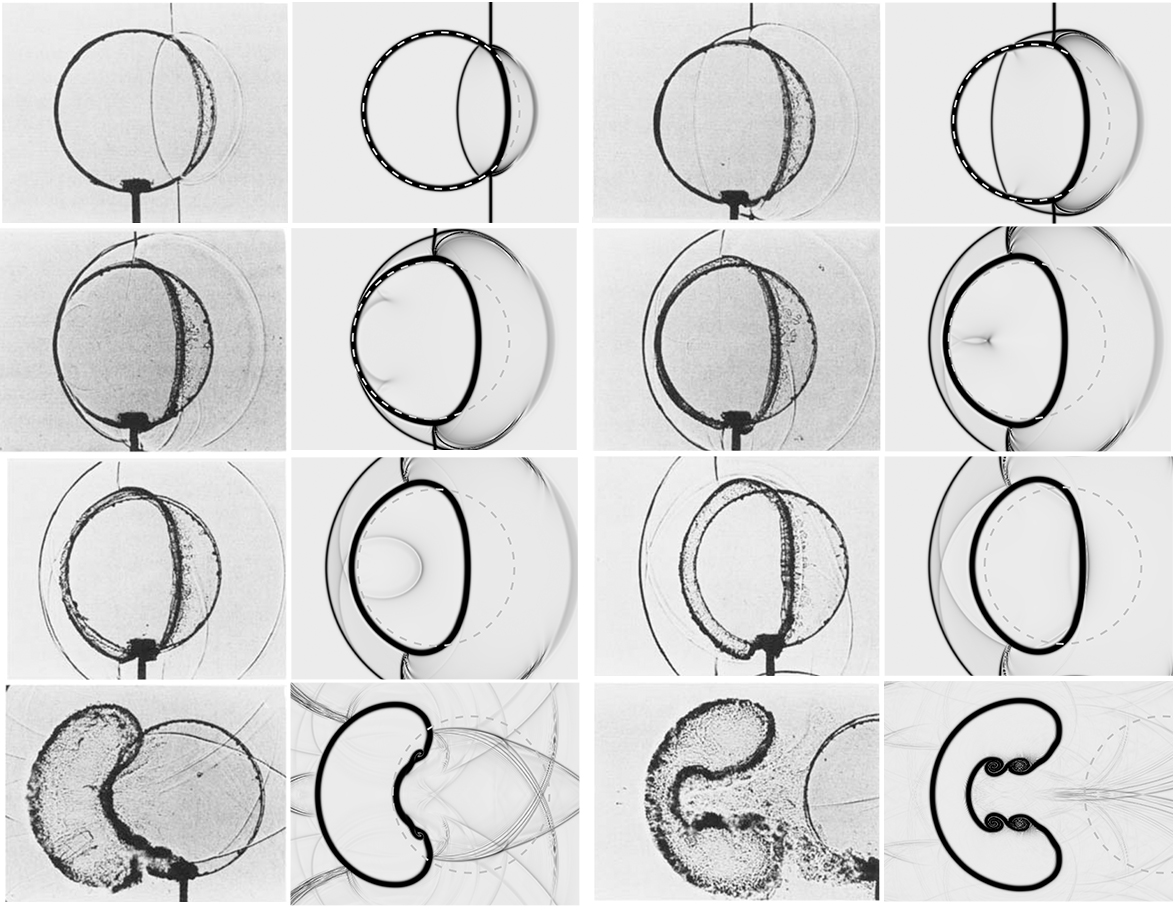}}
			\put(0.09,4.50){(a)}
			\put(3.13,4.50){(b)}
			\put(0.09,3.35){(c)}
			\put(3.13,3.35){(d)}
			\put(0.09,2.18){(e)}
			\put(3.13,2.18){(f)}
			\put(0.09,1.02){(g)}
			\put(3.13,1.02){(h)}
		\end{picture}
		\caption{Schlieren snapshots of the interaction of a planar $M_s=1.22$ shock in air with a cylindrical Helium
			bubble contaminated with 28\% air. The right frame of each subfigure shows the numerical schlieren of the
			observable simulations, while the left frames show the experimental snapshots from
			\citet{SturtevantB:87a}. The snapshots are taken at times (a) 22 $\mu s$ (32 $\mu s$), (b) 42 $\mu s$ (52
			$\mu s$), (c) 56 $\mu s$ (62 $\mu s$), (d) 72 $\mu s$ (72 $\mu s$), (e) 82 $\mu s$ (82 $\mu s$), (f) 102
			$\mu s$ (102 $\mu s$), (g) 245 $\mu s$ (245 $\mu s$), and (h) 427 $\mu s$ (427 $\mu s$). The times reported in
			parentheses are the corresponding reported times in the experiment. For the observable simulation, a
			resolution of $\Delta x = 111\, \mu m$ and a non-dimensional observability limit of $\alpha/\Delta x = 2$ are
			used. The dashed line in the numerical schlieren and solid circular line in experimental ones are the initial
			position of the bubble.}
		\label{Fig:2DShock_He_Bubble}
	\end{center}
\end{figure}

{\Cref{Fig:2DShock_He_Bubble}} shows a comparison of the numerical schlieren snapshots of our simulation
with the schlieren images from the experiment \cite{SturtevantB:87a}.
In this problem the shock meets the interface at a zero angle at $t=0\,\mu s$, which results in a regular
refraction that initially consists of a reflected rarefaction wave and a transmitted shock wave, see snapshot (a). As
the incident shock moves downstream, the angle between the shock and interface increases and the reflected wave first
turns into a Mach wave and then a shock wave before the refraction transforms into an irregular one. In the
irregular refraction, the refracted shock runs ahead of the incident shock and a side shock wave appears, which creates
a Mach stem that can be seen clearly in snapshots (b) and (c). For more details on the regular and irregular shock
refraction in a slow-fast configuration, the reader is referred to
\cite{HendersonL:78a,HendersonL:91a,TheofanousT:05a}. The refracted shock wave also gets reflected internally, where
the shock touches the interface, see snapshots (b) and (c). The numerical schlieren results shown in snapshots (a), (b),
and (c) are processed at instants, which more closely resemble the experimental snapshots. As mentioned in the caption
of the figure, the timing of these snapshots are slightly different from the corresponding times reported in the
experiment. This mismatch is also reported in the work of other authors \cite{Colonius:14a,MaoD:13a,MuletP:03a}, which
might be associated with the timing error in the experiment or the non-uniform contamination which is not accounted for
in the simulations. At $t=56\, \mu s$ the refracted shock retransmits into the air, see snapshot (d). The internally
reflected shock wave partly transmits and runs behind the doubly transmitted shock wave and partly reflects upstream
(circular shape inside the bubble), see snapshot (e). At $t=102\,\mu s$, snapshot (f), the internally reflected shock
wave is transmitted into the air and surrounds the bubble. In snapshot (g), the beginning of interface roll up is
visible at the upstream side of the bubble; at this time the reflected wave from the top and bottom wall have already
interacted with the bubble and can be seen in the front and back of the bubble. The vorticity generated at the interface
by the passage of the shockwave creates a jet on the upstream side of the bubble, which evolves the bubble into a
kidney-like shape, see snapshot (h). The comparisons in this figure show a good agreement between the interface shape
and the wave locations in the observable results with the reported experiment results.

\begin{figure}[htbp]
	\begin{center}
		\setlength{\unitlength}{0.178\linewidth}
		\stackinset{l}{0.09in}{t}{-0.09in}{\footnotesize Experiment \nolink{\cite{SturtevantB:87a}}}{%
			\stackinset{l}{0.09in}{t}{0.07in}{\footnotesize (a)}{\includegraphics[height=\unitlength,trim={0in 0in 0in 0mm},clip]{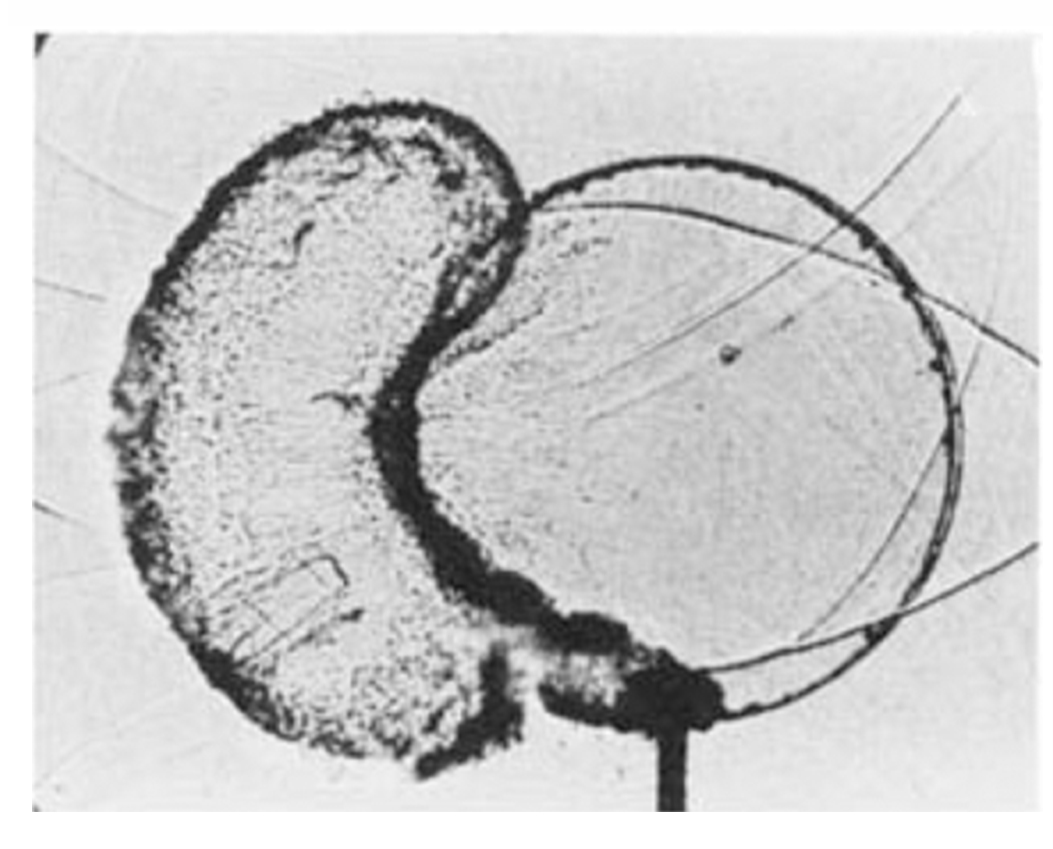}}}
		\stackinset{r}{0.08in}{b}{0.05in}{\footnotesize $\Delta x=56\,\mu m$}{%
			\stackinset{l}{0.06in}{t}{-0.09in}{\footnotesize Computation \nolink{\cite{KarniS:96a}}}{%
				\stackinset{l}{0.06in}{t}{0.07in}{\footnotesize (b)}{\includegraphics[height=\unitlength,trim={0in 0in 0in 0mm},clip]{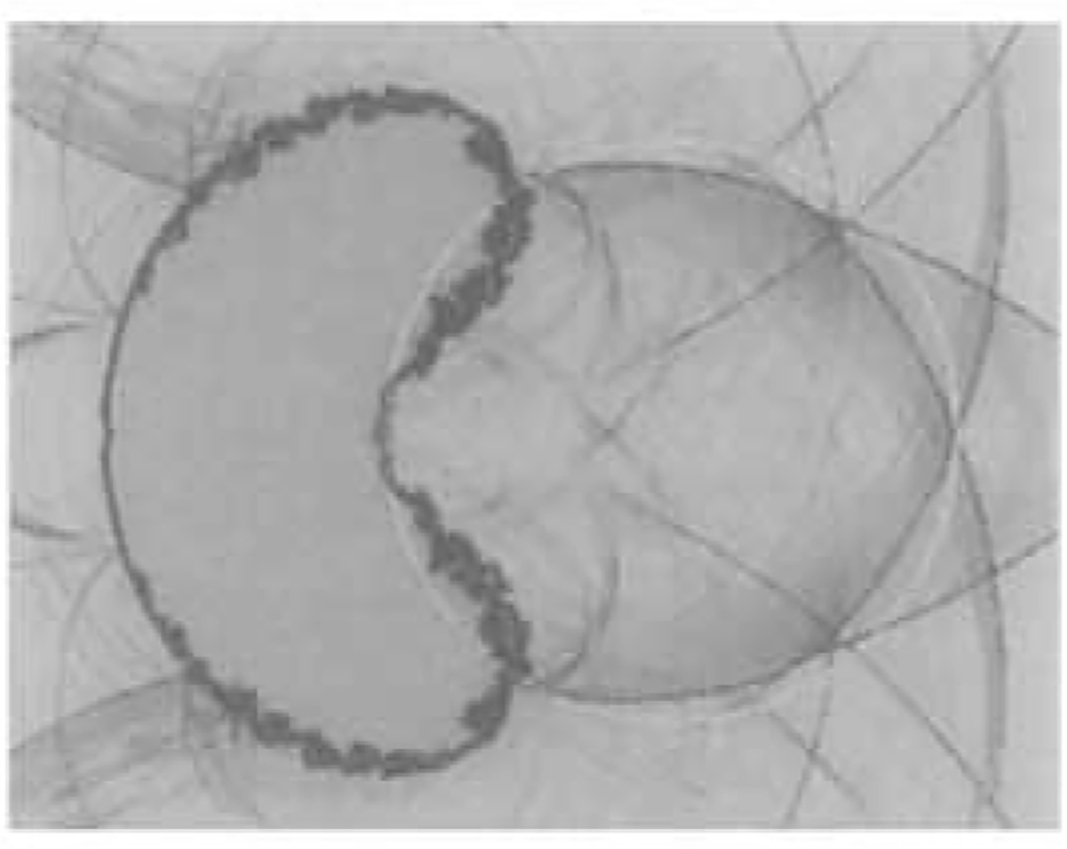}}}}
		\stackinset{r}{0.09in}{b}{0.05in}{\footnotesize $\Delta x=111\,\mu m$}{%
			\stackinset{l}{0.06in}{t}{-0.09in}{\footnotesize Computation, Current Result}{%
				\stackinset{l}{0.06in}{t}{0.07in}{\footnotesize (c)}{\includegraphics[height=\unitlength,trim={0in 0in 0in 0mm},clip]{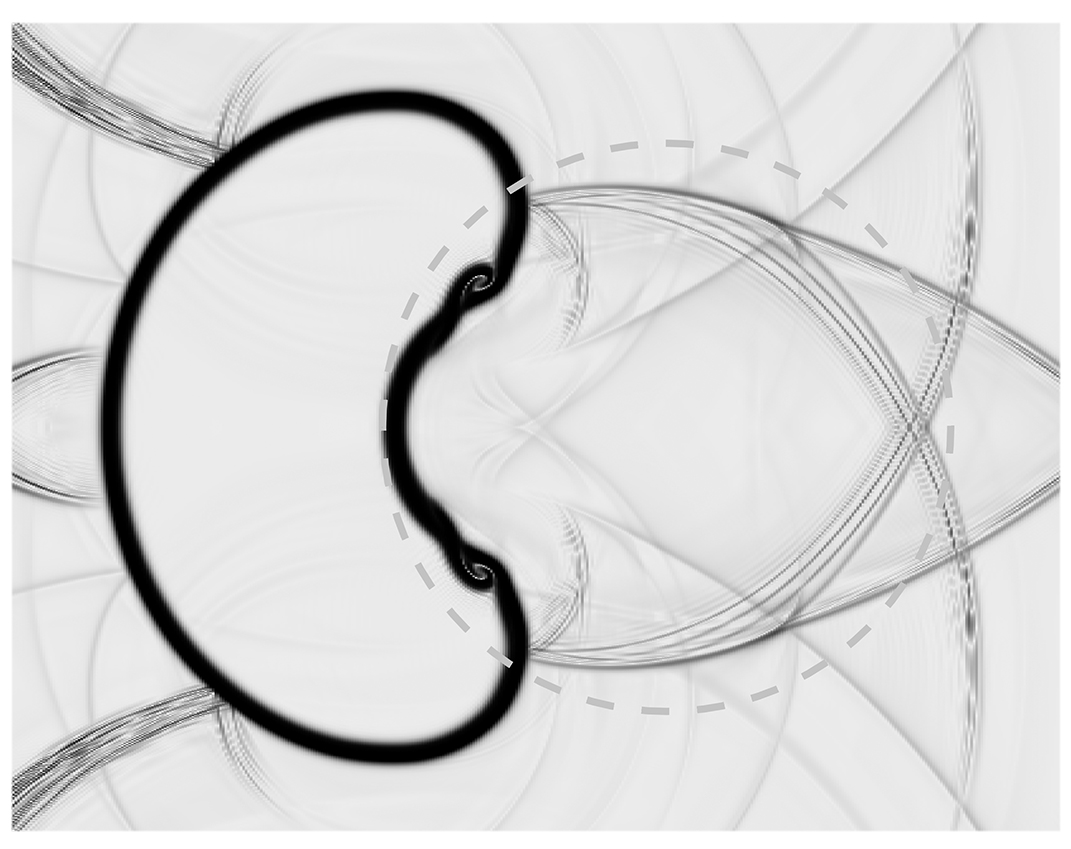}}}}
		\stackinset{r}{0.08in}{b}{0.05in}{\footnotesize $\Delta x=50\,\mu m$}{%
			\stackinset{l}{0.06in}{t}{-0.09in}{\footnotesize Computation \nolink{\cite{Colonius:14a}}}{%
				\stackinset{l}{0.06in}{t}{0.07in}{\footnotesize (d)}{\includegraphics[height=\unitlength,trim={0in 0in 0in 0mm},clip]{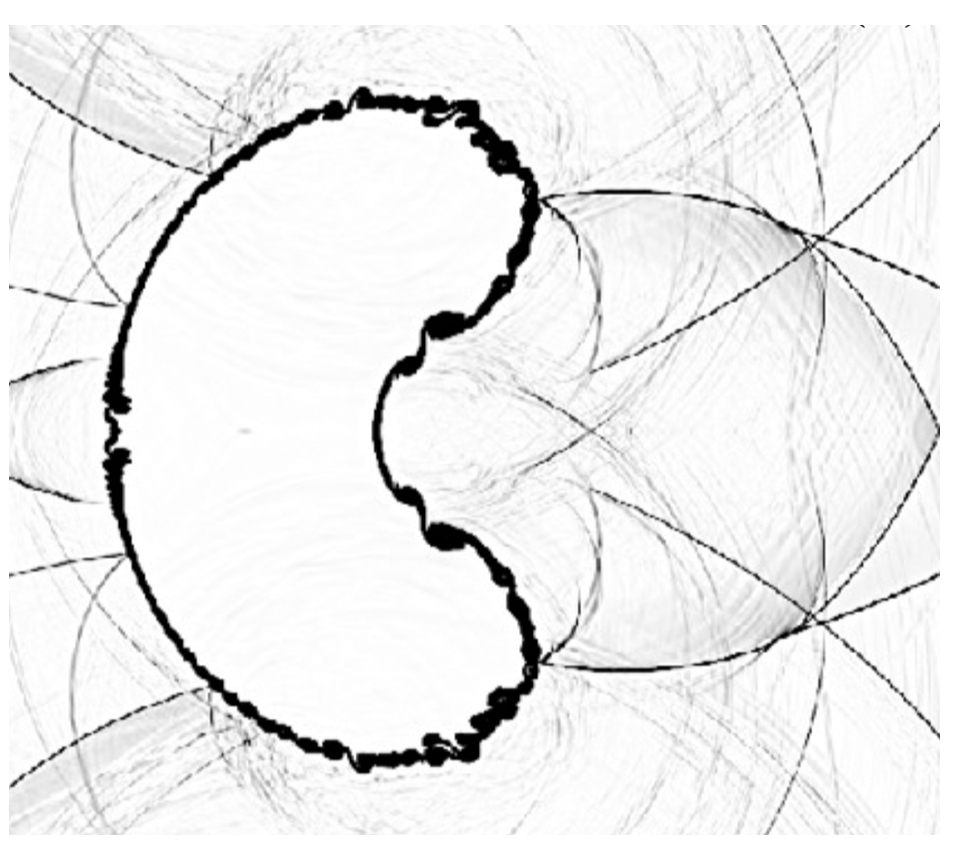}}}}
		\stackinset{r}{0.08in}{b}{0.05in}{\footnotesize $\Delta x=6\,\mu m$}{%
			\stackinset{l}{0.0in}{t}{-0.09in}{\scriptsize Computation \nolink{\cite{KoumoutsakosP:10a}}}{%
				\stackinset{l}{0.05in}{t}{0.07in}{\textcolor{white}{\footnotesize (e)}}{\includegraphics[height=\unitlength,trim={0in 0in 0in 0mm},clip]{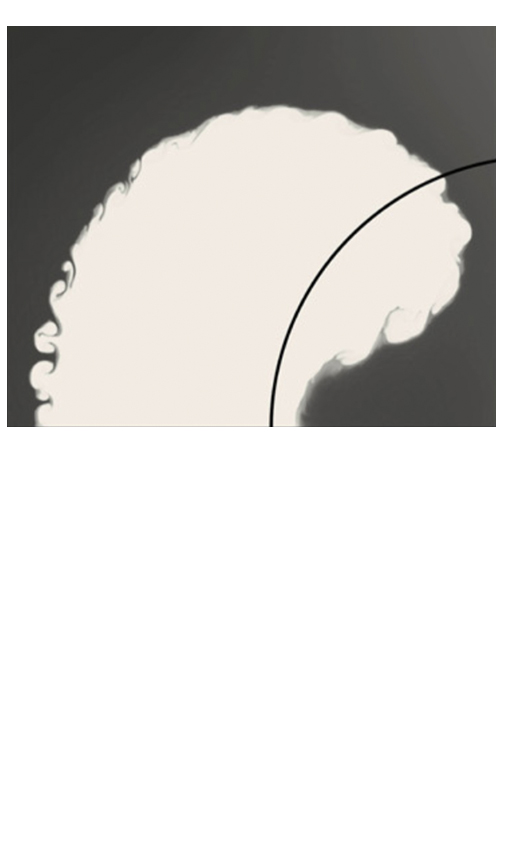}}}}

		\caption{Effect of Euler's infinite resolution on the numerical results. (a) experimental schlieren at
			$t=245\,\mu s$ from \citet{SturtevantB:87a} for shock-Helium+28\%Air bubble problem shown in
				{\cref{Fig:2DShock_He_Bubble}}. (b), (c), and (d) are numerical schlieren images from the simulations by
			\citet{KarniS:96a}, the observable method (this paper), and \citet{Colonius:14a}, respectively. (e) is the
			density contour for a shock-Helium simulation very similar to the simulation reported here without the air
			contamination inside the bubble computed by {\citet{KoumoutsakosP:10a}}. The computational resolution of each
			simulation is mentioned on the bottom of its snapshot.  }
		\label{Fig:effectOfNumericAndGridResolutionOnEuler}
	\end{center}
\end{figure}

In the set of two-phase Euler equations, the thickness of the interface is theoretically zero. Unless the interface is
tracked exactly, as opposed to captured, it is not possible to simulate a zero thickness with any numerical method, this
means that it is the numerical method and computational resolution that dictates the smallest scales of the flow close
to interface. As a result, most numerical methods cannot have a grid-independent solution, especially when there are
small scale features growing next to the interface.  {\cref{Fig:effectOfNumericAndGridResolutionOnEuler}} shows the
numerical schlieren for the shock-bubble problem described above at $t=245\,\mu s$. Snapshot (a) is the experimental
schlieren while snapshot (d) is the result from our method. Snapshot (b) is the numerical schlieren from
\citet{KarniS:96a}, which is achieved by solving the two-phase Euler equation in non-conservative form using an
advection equation for the mass fraction, with a computational resolution $\Delta x=56\,\mu m$. Snapshot (d) is the
result computed by \citet{Colonius:14a} using a fifth order finite volume WENO method accompanied with a volume
fraction equation for capturing the interface. Snapshot (e) is a slightly different simulation in that the bubble is not
contaminated with air (pure Helium bubble). It is computed by \citet{KoumoutsakosP:10a} using a fifth order finite
volume WENO method with an advection equation for a smoothed Heaviside function for capturing the interface, included
here because of its high spatial resolution. The observable result, (c), more closely resembles the experiment. The
three other numerical results show very different interfacial instability depending on the numerical method and the
computational resolution. Snapshot (e) with the highest resolution shows that as the resolution is increased the
interface instability starts growing much earlier and consequently results in different outcome at a later
time. Obviously, by adding viscous terms, and solving the Navier-Stokes equations, some of these issues can be
subdued. However, it will add considerable computational cost that might not be nessessary.

\subsection{Effect of observability limit and its proper choice}
\begin{figure}[htbp]
	\def\stackalignment{l}
	\begin{center}
		\stackinset{c}{0.0in}{b}{0.030in}{\footnotesize$\Delta x = 222\, \mu m, \alpha = 222\, \mu m$}{%
			\stackinset{l}{0.025in}{t}{0.025in}{\footnotesize (a)}{\includegraphics[width=0.20\linewidth,trim={80mm 0mm 170mm 0mm},clip]{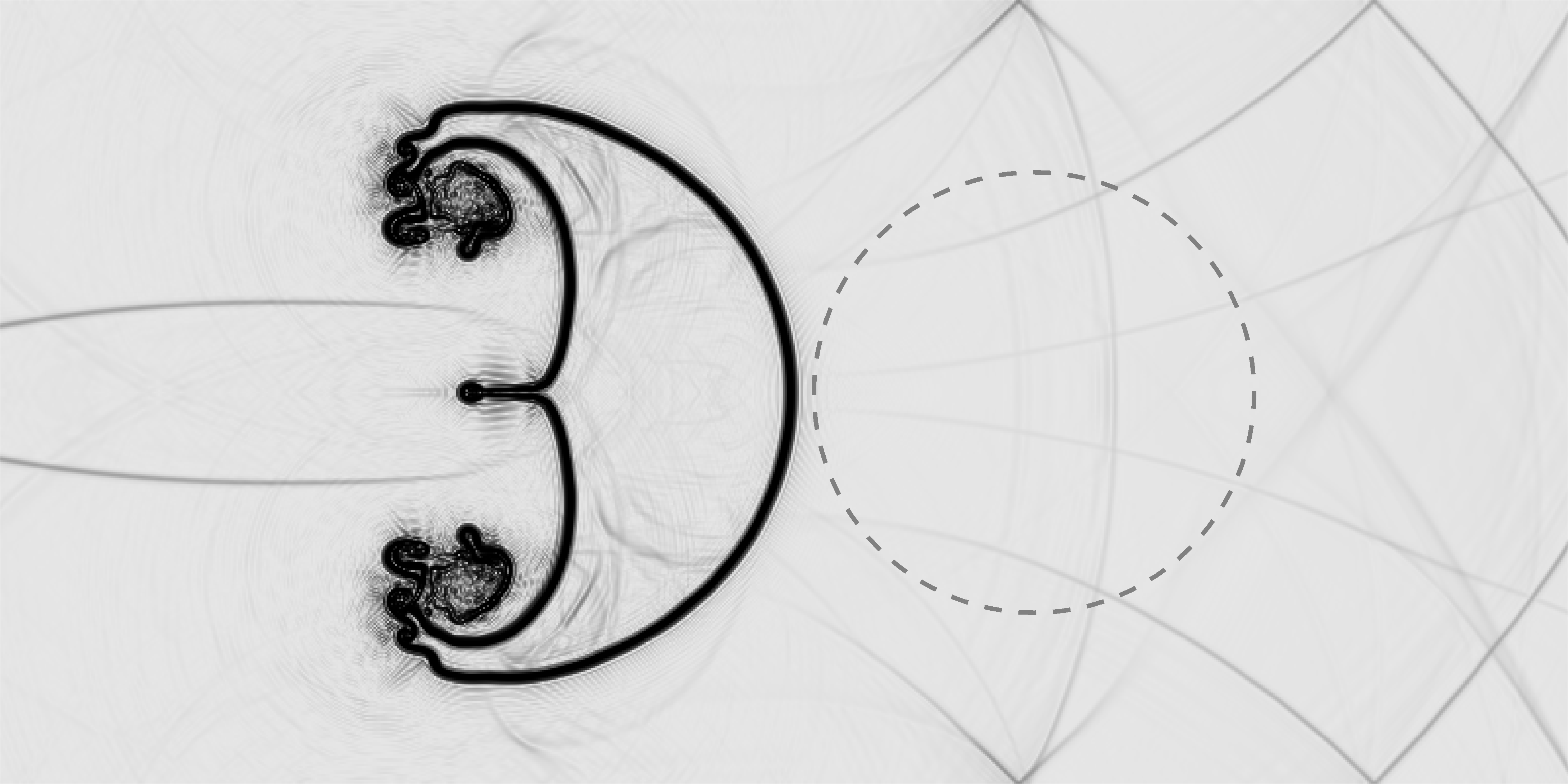}}}
		\stackinset{c}{0.0in}{b}{0.030in}{\footnotesize$\Delta x = 222\, \mu m, \alpha = 444\, \mu m$}{%
			\stackinset{l}{0.025in}{t}{0.025in}{\footnotesize (b)}{\includegraphics[width=0.20\linewidth,trim={80mm 0mm 170mm 0mm},clip]{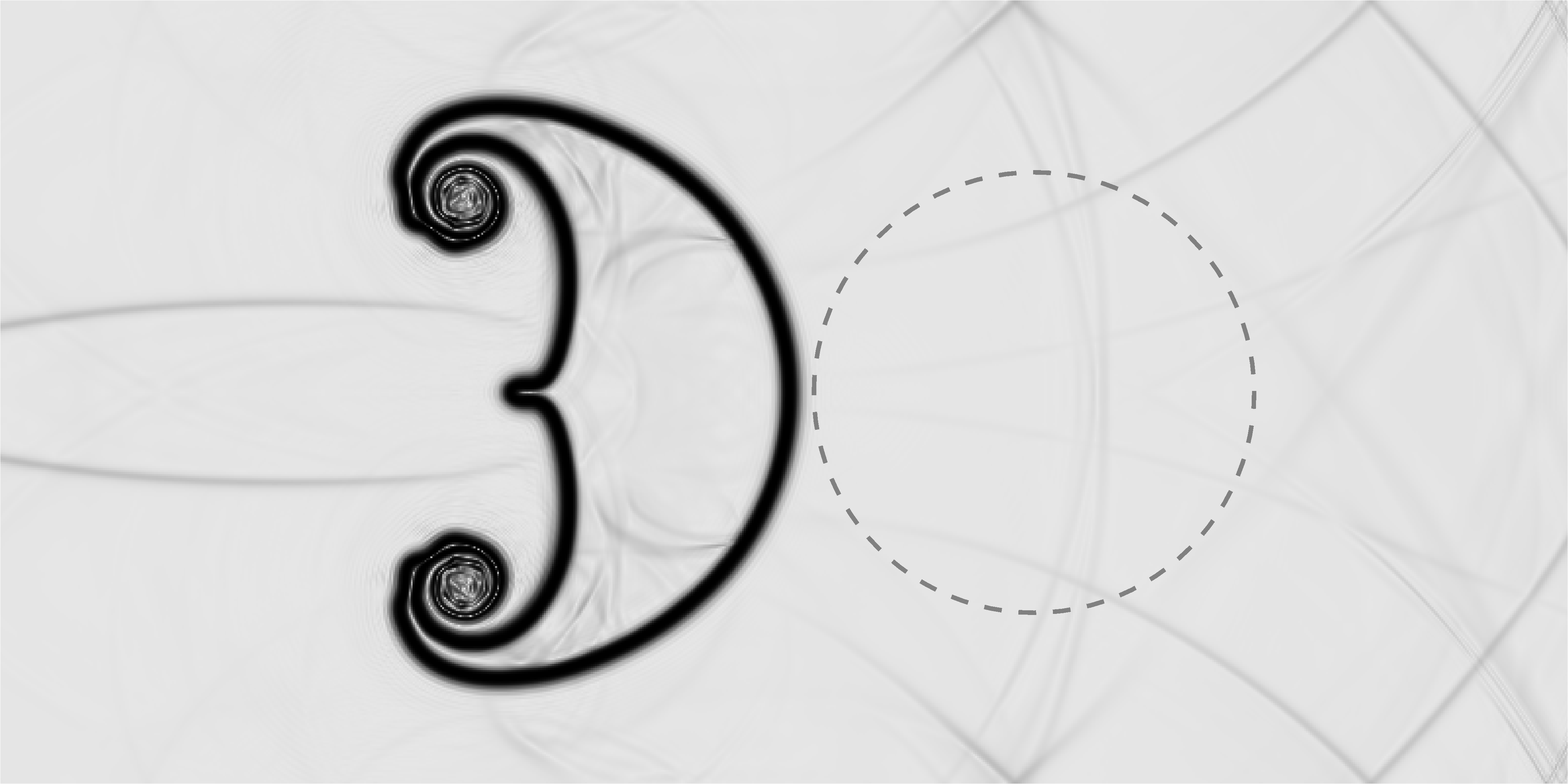}}}
		\stackinset{c}{0.0in}{b}{0.030in}{\footnotesize$\Delta x = 222\, \mu m, \alpha = 666\, \mu m$}{%
			\stackinset{l}{0.025in}{t}{0.025in}{\footnotesize (c)}{\includegraphics[width=0.20\linewidth,trim={80mm 0mm 170mm 0mm},clip]{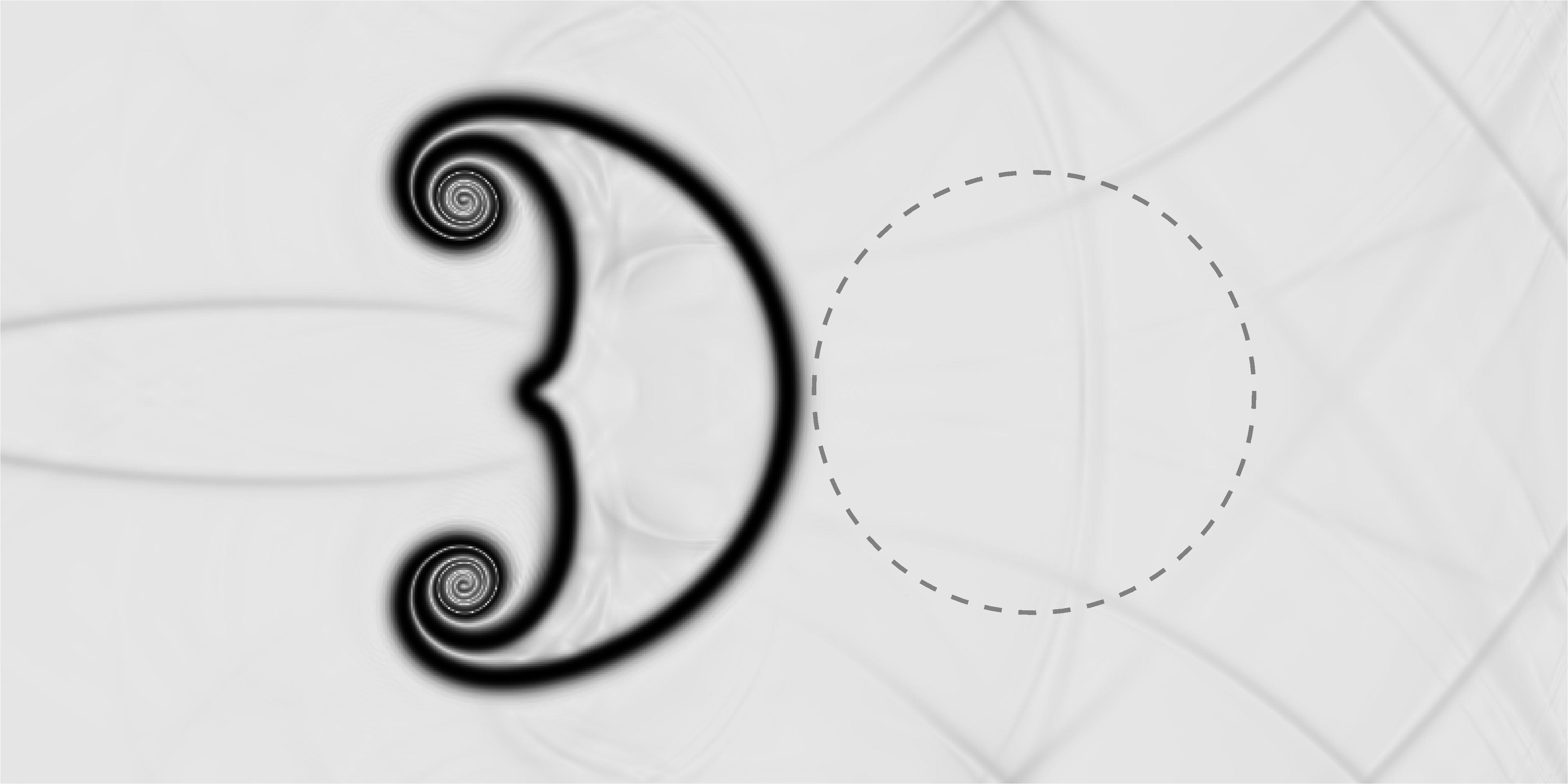}}}
		\stackinset{c}{0.0in}{b}{0.030in}{\footnotesize$\Delta x = 444\, \mu m, \alpha = 444\, \mu m$}{%
			\stackinset{l}{0.025in}{t}{0.025in}{\footnotesize (d)}{\includegraphics[width=0.20\linewidth,trim={80mm 0mm 170mm 0mm},clip]{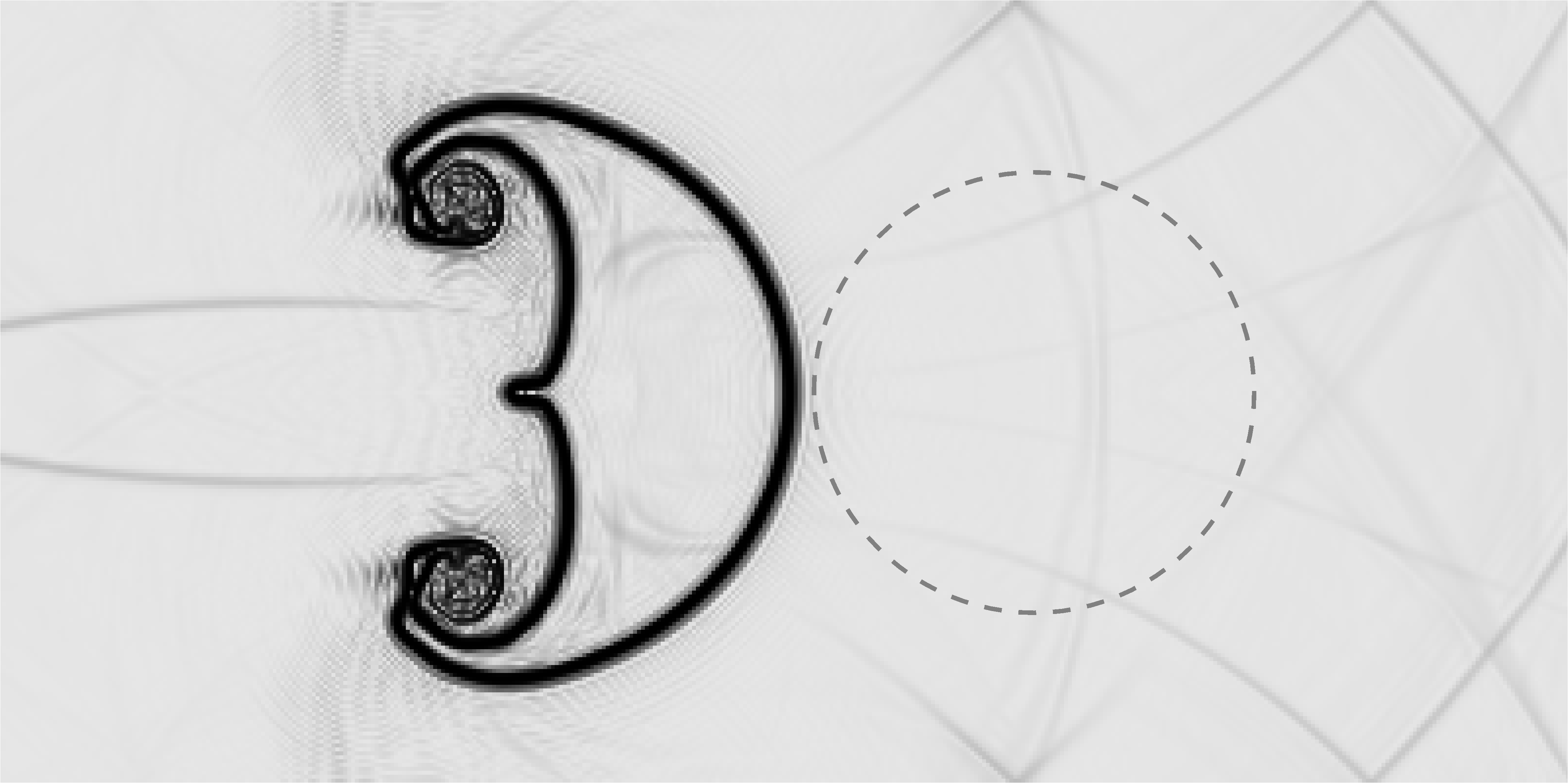}}}
		\caption{The observability limit choice and its effect. Numerical schlieren images shock-R22
			bubble are shown at time $694\, \mu s$. (a), (b), and (c) are computed using grid resolution
			$\Delta x = 222\, \mu m$ while (d) is obtained using $\Delta x = 444\, \mu m$. The observabilty
			limits are (a) $\alpha = 222\, \mu m$ or $\alpha/\Delta x = 1$, (b) $\alpha = 444\, \mu m$ or
			$\alpha/\Delta x = 2$, (c) $\alpha = 666\, \mu m$ or $\alpha/\Delta x = 3$, and (d)
			$\alpha = 444\, \mu m$ or $\alpha/\Delta x = 1$. Dashed line shows the initial location of the
			bubble.}
		\label{Fig:shock_R22_bubble_effect_of_observability}
	\end{center}
\end{figure}

The observability limit specifies how much of high wave numbers or details of the flow will be captured by the
set of observable equations. Intuitively, the observability limit needs to be of the order of grid resolution. For
moderate shock-tube problems, the minimum non-dimensional observability limit from our experience can be as low as 0.5
($\alpha=0.5\Delta x$); note that the length scales below $\Delta x$ are not resolved by the limit of computational
resolution. However, as the strength of the shock increases and more complexities like interfaces are added to the
problem, the minimum acceptable non-dimensional observability limit increases. For the Helmholtz kernel used here,
usually a non-dimensional observability limit of slightly higher than one is recommended. Obviously, as the
observability limit increases, the equations become more regularized and less susceptible to instabilities with the
caveat that some of the small flow structures and details will not be
captured. \Cref{Fig:shock_R22_bubble_effect_of_observability} shows an example of this trend demonstrated by solving the
same shock-R22 bubble interaction problem for different observability limits. As shown in part (a), (b), and (c) of the
figure, by increasing the observability limit, but keeping the same numerical resolution some of the details of the flow
are lost. Part (d) of the figure shows the same simulation on a grid twice as coarse as the ones shown in parts (a), (b)
and (c) with an observability limit equal to the one used in (b). It can be seen that the results shown in (b) and (d)
are almost identical, which demonstrates that while we can increase the observability limit as much as we want, any
non-dimensional observability limit that is not as much as possible close to unity will waste computational resources.

\section{Conclusion}
In this work we used the concepts of observability and observable divergence \cite{Mohseni:10w} to develop a method for
simulation of two-phase compressible flows. This method uses an observability limit that represents the length scale
below which flow structures cannot be resolved. The introduction of the limit is applied at the level of the governing
differential equation and therefore is not a filter on an already-computed solution.

An observable equation for tracking the material interface is developed that preserves pressure equilibrium at the
interface. A pseudo-spectral method is used to avoid numerical dissipation. Using no special numerical treatment for
performing high order calculation, e.g. WENO/ENO, and having no numerical dissipation in the computations shows that it
is the governing equations that regularize the evident discontinuities in the problem. A few 1D test cases are studied
and the results are compared with some of the best shock and interface capturing methods in the literature and the exact
Euler solution. In all cases the current method produce comparable or better results. Additionally, two 2D shock-bubble
interactions are studied and the results are compared with available experimental data. The method captured the major
wave and interface speeds correctly and the qualitative comparison of numerical schlieren snapshots show good agreement
with the experiment. We also compared different results from the literature for the shock-Helium bubble interaction and
discussed that because of the infinitely thin interface assumption in the two-phase Euler equations, it is the numerical
method and the computational resolution which dictate the thickness of the interface. As a result, depending on the
numerical method, different results may be achieved.
The observable
approach, by introducing the observability limit, defines the smallest length scales in the flow
This method reduces the computational cost
by approximately one order of magnitude in 2D computations in comparison to other computations and we expect that the
savings would be considerably more in 3D. Finally, we also looked at the effect of the observability limit and
demonstrated the best practices for selecting the optimum observability limit parameter. Here, an interface-capturing
approach is chosen for recognizing the location of the interface since the simplicity of its implementation attracted
attention in fluid dynamics community. However, it is necessary to note that the concept of observability can be
implemented using other interface-tracking approaches. Since the observable method is regularizing the equations at
the level of the PDE, one can use any numerical method to solve the equations and consequently numerical methods for
unstructured grids can be used to deal with complex geometries.

\section*{Acknowledgment}
We thank Dr. Doug Lipinski for his early work on similar problems which helped development of current work. We
would also like to thank Dr. Adam DeVoria for his helpful comments in preparation of this manuscript. This work was
partially supported in part by the Air Force Office of Scientific Research as well as the National Science Foundation.
\pagebreak
\begin{appendix}
	\section{}
	Since the equation of state is used in the derivation of the governing equation for the volume fraction,
	changing the equation of state might result in slight changes in the volume fraction equation. Here, the derivation
	of observable volume fraction equation is presented for a more general equation of state, the stiffened equation of
	state. In this case the derived equation is the same as the one derived for ideal gas equation of state.

		The stiffened equation of state can be written as
	\begin{equation}\rho e = \Gamma p + \Pi ,\quad \,\quad \,\quad \,\quad \,\quad \,\Gamma  = \frac{1}{{\gamma  - 1}},\,\,\,\,\Pi  = \frac{{\gamma {p^\infty }}}{{\gamma  - 1}}
		\label{eqn:StiffenedEOS}\end{equation}
	in which $P^\infty$ is usually zero for gases and non-zero for liquids. To derive the observable volume fraction
	equation, we insert \mbox{\cref{eqn:StiffenedEOS}} into \mbox{\cref{eqn:ObsIntEnergyForInterface}} and by
	rearranging we have:
	\begin{equation}\Gamma \left( {\frac{{\partial p}}{{\partial t}} + \bar u\frac{{\partial p}}{{\partial x}}} \right) + p\left( {\frac{{\partial \Gamma }}{{\partial t}} + \bar u\frac{{\partial \Gamma }}{{\partial x}}} \right) + \left( {\frac{{\partial \Pi }}{{\partial t}} + \bar u\frac{{\partial \Pi }}{{\partial x}}} \right) = 0.\end{equation}
	With the pressure equilibrium requirement, the first parenthesis is zero. Since this equation needs to be
		satisfied for any pressure, the terms in the second and third parentheses should be zero. By defining the
		mixture property $\Pi$ similar to \mbox{\cref{eqn:mixturematerialprop}}, $\Pi  = \frac{{\gamma {p^\infty
		}}}{{\gamma  - 1}} = \frac{{{\gamma _1}p_1^\infty {z_1}}}{{{\gamma _1} - 1}} + \frac{{{\gamma _2}p_2^\infty
		{z_2}}}{{{\gamma _2} - 1}}$, and additionally using \mbox{\cref{eqn:mixturematerialprop}}, the simplified
		observable equation for the volume fraction becomes \mbox{\cref{Eqn:ObservableVolumeFraction}}, $\frac{{\partial
		{z _1}}}{{\partial t}} + \overline {\bf{u}} \cdot \nabla {z _1} = 0.$.

\begin{figure}[htp]
		\begin{center}
			\includegraphics[width=0.9\textwidth]{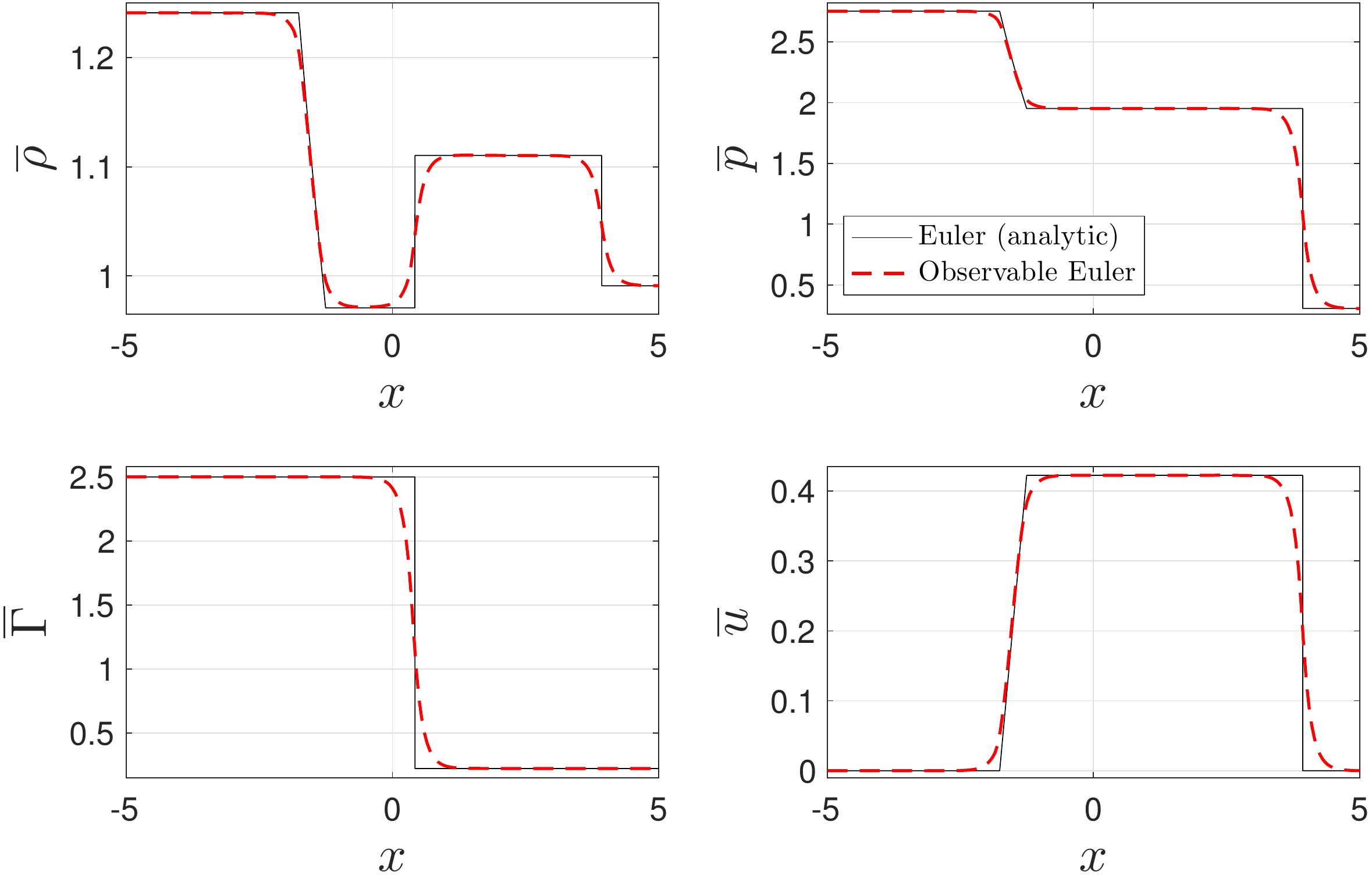}
			\caption{Solution of a gas-liquid shock-tube problem with the initial condition given in \mbox{\Cref{eqn:inital-gas-liquid-shock-tube}}. }
			\label{Fig:Gas-liquid-Shock-tube}
		\end{center}
	\end{figure}

	Here, the observable equations are tested using a gas-liquid Riemann problem used to model underwater explosions
	\cite{CocchiJ:96a, ShyueKM:98a, Colonius:06a}. The location of the discontinuity is at the origin at time equal zero
	and the initial condition is:
	\begin{equation}
		\begin{gathered}
			(\rho {\text{,}}u{\text{,}}P{\text{,}}\gamma {\text{,}}{P_\infty })_{\text{L}}^{\text{T}} = {(1.241{\text{,}}0{\text{,}}2.753{\text{,}}1.4{\text{,}}0)^{\text{T}}}{\text{,}} \hfill \\
			(\rho {\text{,}}u{\text{,}}P{\text{,}}\gamma {\text{,}}{P_\infty })_{\text{R}}^{\text{T}} = {(0.991{\text{,}}0{\text{,}}3.059 \times {10^{ - 4}}{\text{,}}5.5{\text{,}}1.505)^{\text{T}}}{\text{.}} \hfill \\
		\end{gathered}
		\label{eqn:inital-gas-liquid-shock-tube}
	\end{equation}
	\mbox{\Cref{Fig:Gas-liquid-Shock-tube}} shows the result of this problem by solving the two-phase observable
	Euler equations with a grid resolution of $\Delta x = 1/20$ or 200 points across physical domain and an
	observability limit $\alpha = \Delta x$. As shown in the figure, the observable method produce good results and
	captures all the wave speeds correctly.

\end{appendix}

\bibliographystyle{unsrtnat} 
\bibliography{RefA2}

\end{document}